%% file: AAmain.tex
\documentclass[aps,prx,twocolumn,reprint,amsmath,amssymb,superscriptaddress,floatfix,footinbib,longbibliography, tikz, dvipsnames]{revtex4-1}
\usepackage{natbib}

\usepackage{amsmath}
\usepackage{amssymb}
\usepackage{amsthm}
\usepackage{physics}

\usepackage{array}
\usepackage{multirow}
\usepackage{enumerate}
\usepackage{bm}

\usepackage{graphicx}
\usepackage{empheq}
\usepackage{float}
\usepackage{tabularx}

\newcolumntype{C}{>{\centering}X}

\usepackage{standalone}
\usepackage{tikz}
\usepackage{pgfplots}
\pgfplotsset{compat=1.5}
\usetikzlibrary{arrows}
\usetikzlibrary{decorations.markings}   
\usetikzlibrary{patterns}
\usetikzlibrary{decorations.pathmorphing}

\usepackage{xcolor}
\usepackage{url}
\usepackage[colorlinks]{hyperref}
\hypersetup{%
        plainpages=true,
        breaklinks=true,
        hypertexnames=false,
        pageanchor=true,
        colorlinks=true,
        linkcolor={MidnightBlue},
        citecolor={RedOrange},
        urlcolor={MidnightBlue},
        anchorcolor={black}
      }
\usepackage[all]{hypcap} 

\usepackage{mleftright} 

\newcommand{\figref}[1]{\mbox{Fig.~\ref{#1}}}
\newcommand{\tabref}[1]{\mbox{Table~\ref{#1}}}
\newcommand{\secref}[1]{\mbox{Sec.~\ref{#1}}}

\newcommand{\appref}[1]{\mbox{Appendix~\ref{#1}}} 
\renewcommand{\eqref}[1]{\mbox{Eq.~(\ref{#1})}}
\newcommand{\eqsref}[2]{\mbox{Eqs.~(\ref{#1})-(\ref{#2})}}
\newcommand{\secsref}[2]{\mbox{Secs.~\ref{#1}-\ref{#2}}}

\newcommand{\figpanel}[2]{Fig.~\hyperref[#1]{\ref*{#1}(#2)}}
\newcommand{\figpanels}[3]{Fig.~\hyperref[#1]{\ref*{#1}(#2)-(#3)}}
\newcommand{\figpanelNoPrefix}[2]{\hyperref[#1]{\ref*{#1}(#2)}}

\newcommand{\nn}{\nonumber}
\newcommand{\Hc}{\text{H.c.}}
\newcommand{\R}{\mathbb{R}}

\newcommand{\D}{\mathcal{D}} 
\newcommand{\COL}{\mathcal{L}} 
\newcommand{\SCA}{\mathcal{S}} 
\newcommand{\Gammacol}{\Gamma_\text{coll}}
\newcommand{\GammaD}{\Gamma_{D_{S/T}}}
\newcommand{\GammaA}{\Gamma_a}
\newcommand{\GammaB}{\Gamma_b}
\newcommand{\omegaA}{\omega_a}
\newcommand{\omegaB}{\omega_b}
\newcommand{\OmegaA}{\Omega_a}
\newcommand{\OmegaB}{\Omega_b}
\newcommand{\deltaA}{\delta_a}
\newcommand{\deltaB}{\delta_b}
\newcommand{\smapb}{{\sigma_-^a\sigma_+^b}}
\newcommand{\spamb}{{\sigma_+^a\sigma_-^b}}
\newcommand{\sza}{{\sigma_z^a}}
\newcommand{\szb}{{\sigma_z^b}}
\newcommand{\sma}{{\sigma_-^a}}
\newcommand{\smb}{{\sigma_-^b}}
\newcommand{\spa}{{\sigma_+^a}}
\newcommand{\spb}{{\sigma_+^b}}

\newcommand{\series}{\;\triangleleft\;}

\AtBeginDocument{%
    \newwrite\bibnotes
    \def\bibnotesext{Notes.bib}
    \immediate\openout\bibnotes=\jobname\bibnotesext
    \immediate\write\bibnotes{@CONTROL{REVTEX41Control}}
    \immediate\write\bibnotes{@CONTROL{%
    apsrev41Control,author="08",editor="1",pages="0",title="0",year="1"}}
     \if@filesw
     \immediate\write\@auxout{\string\citation{apsrev41Control}}%
    \fi
}%

\begin{document}

\title{Chiral quantum optics with giant atoms}
\date{\today}

\author{Ariadna Soro}
\email{soro@chalmers.se}
\affiliation{Department of Microtechnology and Nanoscience, Chalmers University of Technology, 412 96 Gothenburg, Sweden}

\author{Anton Frisk Kockum}
\email{anton.frisk.kockum@chalmers.se}
\affiliation{Department of Microtechnology and Nanoscience, Chalmers University of Technology, 412 96 Gothenburg, Sweden}

\begin{abstract}
In quantum optics, it is common to assume that atoms are point-like objects compared to the wavelength of the electromagnetic field they interact with. However, this dipole approximation is not always valid, e.g., if atoms couple to the field at multiple discrete points. Previous work has shown that superconducting qubits coupled to a one-dimensional waveguide can behave as such ``giant atoms'' and then can interact through the waveguide without decohering, a phenomenon that is not possible with small atoms. Here, we show that this decoherence-free interaction is also possible when the coupling to the waveguide is \textit{chiral}, i.e., when the coupling depends on the propagation direction of the light. Furthermore, we derive conditions under which the giant atoms in such chiral architectures exhibit dark states. In particular, we show that unlike small atoms, giant atoms in a chiral waveguide can reach a dark state even without being excited by a coherent drive. We also show that in the driven-dissipative regime, dark states can be populated faster in giant atoms. The results presented here lay a foundation for applications based on giant atoms in quantum simulations and quantum networks with chiral settings.
\end{abstract}

\maketitle

\section{Introduction}

In recent years, new paradigms in quantum optics have emerged. On one hand, it has been shown that the dipole approximation, i.e., the assumption that atoms are \textit{small} compared to the wavelength of the light they interact with, is not always valid.
We typically refer to atoms that break this approximation as \emph{giant}, since they can couple to light -- or other bosonic fields -- at several points, which may be spaced wavelengths apart.
The physics of such atoms has mostly been studied from a theoretical perspective~\cite{Kockum_Review, Kockum14, Guo17, Kockum18, GonzalezTudela19, Ask2019, Guimond20, Guo20, Guo2020a, Zhao2020, Cilluffo2020, Carollo2020, Longhi2020, Ask20, Wang21, Cheng2021, Du2021, Du2021a, Vega2021}, with findings including a frequency-dependent relaxation rate~\cite{Kockum14} and decoherence-free interaction between multiple giant atoms coupled to a waveguide~\cite{Kockum18}.
Experimental demonstrations of giant atoms have also been realized, by coupling superconducting artificial atoms~\cite{Gu2017, Kockum2019a, Blais2020} to surface acoustic waves~\cite{Gustafsson14, Aref2016, Manenti2017, Noguchi2017, Satzinger2018, Moores2018, Bolgar2018, Sletten2019, Bienfait2019, Andersson2019, Bienfait2020, Andersson2020} and microwave waveguides~\cite{Kannan20, Vadiraj21}.

Another new approach in quantum optics is based on chiral interfaces, where the coupling between light and matter depends on the propagation direction of the light~\cite{Lodahl17, Scheucher20}.
Such chirality emerges naturally in optical nanofibers when light is strongly transversely confined~\cite{Mitsch14, Petersen14, BliokhNori15} and it is also achievable in microwave waveguides by using circulators~\cite{Gu2017, circulators14, circulators15, circulators17, circulators18}, sawtooth lattices~\cite{SanchezBurillo20}, or entangled states between quantum emitters~\cite{Gheeraert20}.
Besides photonic reservoirs, other architectures have been proposed to realize chiral coupling, such as phononic waveguides consisting of Bose-Einstein quasicondensates~\cite{SO11, SO13, Ramos14} or trapped ions~\cite{Vermersch16}, and magnonic waveguides made of spin chains~\cite{Vermersch16, Ramos16}. Chiral quantum optics is increasingly attracting interest~\cite{Stannigel12, GonzalezBallestero15, Pichler15, Guimond16, Bello19, Ozawa19, Scheucher20, Mahmoodian20, Kim21} since it can be harnessed in quantum information for quantum state transfer between qubits and for manipulation of stabilizer codes for quantum error correction~\cite{Lodahl17, Guimond20}.

In this paper, we study the overlap between the aforementioned paradigms in quantum optics, which has barely been explored due to its recentness~\cite{Guimond20, Cilluffo2020, Carollo2020, Wang21, Cheng2021}. In particular, Refs.~\cite{Guimond20, Wang21, Cheng2021} use giant atoms to induce effective chirality and Refs.~\cite{Cilluffo2020, Carollo2020} use a collision model for exploring chiral setups. 
In this manuscript, however, we analyze the behaviour of multiple giant atoms chirally coupled to a one-dimensional (1D) bosonic waveguide through a master-equation treatment. We show that a certain arrangement of the connection points of the giant atoms allows them to interact without decohering into the waveguide, in the same way they do when the coupling is bidirectional~\cite{Kockum18, Kannan20}.
This protected interaction is independent of the state of the giant atoms and not accessible to small atoms.

Furthermore, we look into dark states~\cite{Dicke54, Lenz93, LeKien05} as another -- less robust -- way of protecting the system against decoherence.
We derive conditions for the existence of such states in undriven atomic ensembles and show that, unlike small atoms~\cite{Pichler15}, certain configurations of giant atoms allow for perfect subradiance regardless of the chirality of their coupling.
We also extend this analysis to the driven-dissipative regime and find that, compared to small atoms~\cite{Pichler15}, giant atoms can populate dark states faster.

The theoretical treatment presented here is valid for any two-level systems that can chirally couple to 1D waveguides of bosonic nature at multiple points.
This implies that numerous architectures could potentially realize this chiral coupling of giant atoms, e.g., transmon qubits~\cite{Koch2007} coupled to meandering transmission lines with circulators, ultracold atoms in dynamical state-dependent optical lattices~\cite{GonzalezTudela19}, and perhaps also cold atoms coupled to optical nanofibers. 

\begin{figure*}[t]
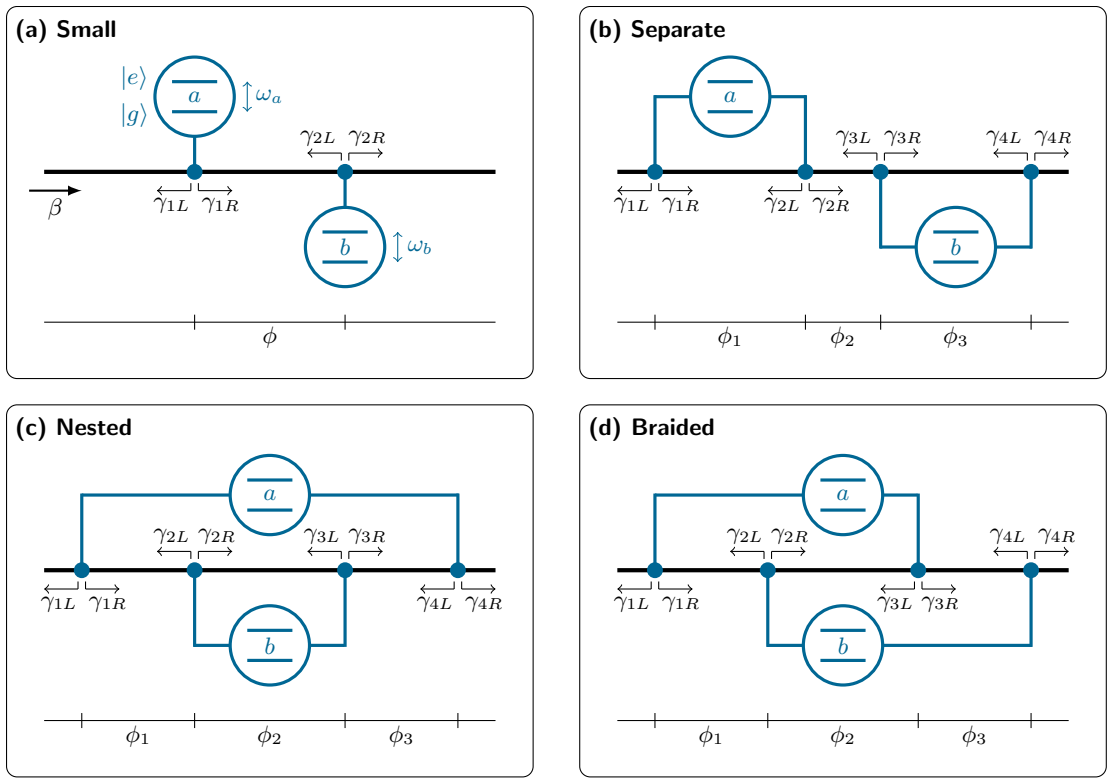

\centering
\includestandalone{Small}
\includestandalone{Separate}
\includestandalone{Nested}
\includestandalone{Braided}
\caption{Diatomic setups considered in the present work. (a) Two small atoms, (b) two separate giant atoms, (c) two nested giant atoms, and (d) two braided giant atoms, coupled to a 1D open waveguide. The atoms are denoted by $a$ and $b$, and their connection points to the transmission line are identified by the relaxation rates $\gamma_k$, for $k=1,2,3,4$. At each coupling point, we distinguish the decay rate to the right- and left-propagating modes as $\gamma_{kR}$ and $\gamma_{kL}$, in such a way that $\gamma_{kR}+\gamma_{kL}=\gamma_k$. The phase shifts between neighbouring coupling points are denoted by $\phi_k$ for $k=1,2,3$.}
\label{fig:setups}
\end{figure*}

The present work can find applications in quantum simulations~\cite{qsim14} of open chiral many-body systems.
In particular, decoherence-free interaction between distant atoms allows the simulation of open quantum many-body systems~\cite{Kannan20} and the results found here extend the simulatability to chiral interfaces.
Furthermore, we believe that giant atoms can be used to generate spatially-entangled photons~\cite{Kannan20b} and entangled cluster states for one-way quantum computing~\cite{Pichler17}. On the other hand, dark states are typically involved in coherent population trapping~\cite{Arimondo76, Alzetta76, Gray78}, electromagnetically induced transparency~\cite{Harris90, Boller91, Harris97, Fleischhauer05, Ask20, Vadiraj21}, and stimulated Raman adiabatic passage~\cite{Bergmann98, Bergmann15, Vitanov17}, which are key phenomena to develop quantum memories~\cite{Fleischhauer02, Lvovsky09, Heshami16} and quantum networks~\cite{Kimble2008, Wehner18}.

This paper is structured as follows.
In \secref{sec:me}, we present a theoretical model that we use to describe two atoms chirally coupled to a 1D open waveguide.
We use the standard master-equation treatment, where the dynamics of the system are obtained by tracing out the waveguide modes in the Born--Markov approximation.
We employ this model to study decoherence-free interaction in \secref{sec:dfi} and find possible dark states in \secref{sec:ds}.
In particular, in \secref{sec:dsu}, we assume that the atoms are not driven, whereas in \secref{sec:dsd} they are driven by a coherent field.
Finally, in \secref{sec:general}, we discuss how the results found generalize to multiple atoms with multiple coupling points, and we conclude in \secref{sec:conclusion} with a summary and an outlook.
We also include two appendices with more detailed derivations of the mathematical model used (\appref{app:me}) and the dark-state conditions (\appref{app:ds}).

\section{Theoretical model}
\label{sec:me}

The setup we initially consider in this article is a system consisting of two two-level quantum emitters (qubits) dissipatively coupled to a bosonic bath, which we assume to be an open 1D waveguide (\figref{fig:setups}).
The two atoms are the simplest layout that allow us to first illustrate the main physical phenomena.
Later, in \secref{sec:general}, we consider a larger system with an arbitrary number of atoms and coupling points, and provide a more comprehensive description of the dynamics.

Depending on the amount of coupling points between the atoms and the waveguide, we distinguish \emph{small} atoms, which are coupled at a single point [\figpanel{fig:setups}{a}], and \emph{giant} atoms, with two or more connection points each [\figpanels{fig:setups}{b}{d}].
For simplicity, we assume here that the giant atoms interact with the waveguide at only two points, and we identify three different topologies according to the arrangement of their coupling points~\cite{Kockum18}: \emph{separate} [\figpanel{fig:setups}{b}], \emph{nested} [\figpanel{fig:setups}{c}], and \emph{braided} [\figpanel{fig:setups}{d}].

Besides the amount and arrangement of the connection points, the coupling of the atoms to the reservoir can also be characterized by the directionality of the propagating modes in the waveguide.
Throughout this article, we say that atoms couple \emph{chirally} to a waveguide when their bare relaxation rate (the relaxation rate before any interference effects are taken into account) is generally different towards the right and left directions, i.e., $\gamma_R\neq\gamma_L$.
Consequently, there are two limiting cases: the \emph{bidirectional} or \emph{nonchiral} case, where atoms couple symmetrically to the right and left ($\gamma_R=\gamma_L$), and the \emph{unidirectional} or \emph{cascaded} case, where atoms couple to modes propagating in only one direction (e.g., $\gamma_L=0$).

We remark that chiral coupling does not in itself break Lorentz reciprocity~\cite{Deak2012} in the waveguide.
For instance, in photonic nanostructures, the strong confinement of light causes a spin-momentum locking effect, which allows directional propagation of photons without breaking reciprocity.
In the model we present here, we consider the bath-system ensemble to be nonreciprocal, whether it consists of a reciprocal bath with chiral coupling to the system or an intrinsically nonreciprocal bath.
This means it does not necessarily apply to unidirectional transducers (UDTs)~\cite{Ekstrom2017, Ekstrom2019} and other reciprocal devices.

Formally, we can derive a master equation for the density matrix $\rho$ of all the setups described above. In order to do so, we assume that the coupling of each atom $j\in\{a,\,b\}$ is weak compared to their transition frequency, i.e., $\Gamma_j \ll \omega_j$, and that the travel time between connection points is negligible compared to the relaxation times $1/\Gamma_j$ of all the atoms.
Then, we use the SLH formalism \cite{SLH1, SLH2, SLH3} for cascaded systems to derive a master equation for the diatomic ensemble. The SLH triplet, which consists of a scattering matrix $\SCA$, a vector $\COL$ of $n$ collapse operators that describe the coupling of the system to the transmission line, and the Hamiltonian $H$ of the system, yields the master equation for the system in the Lindblad form ($\hbar = 1$ throughout this article):
\begin{equation}
    \dot{\rho} = -i[H, \rho] + \sum_{k=1}^n   \D[\COL_k]\rho,
    \label{eq:me_Lindblad}
\end{equation}
where $\D[X] \rho = X\rho X^\dag-\frac{1}{2}\acomm{X^\dag X}{\rho}$ are Lindblad superoperators.
In particular, as shown in \appref{app:me}, we can write the master equation for all the setups in \figref{fig:setups} as follows:
\begin{equation}
\begin{aligned}
	\dot{\rho} = &
	-i \comm{\omegaA'\frac{\sza}{2} + \omegaB'\frac{\szb}{2} + \mleft(g\smapb +
	\Hc\mright) + H_d}{\rho}\\
	&+ \GammaA \D[\sma]\rho + \GammaB \D[\smb]\rho \\
	& +\mleft[\Gammacol \mleft(\sma\rho\spb-\frac{1}{2}\acomm{\smapb}{\rho}\mright) + \Hc \mright],
	\label{eq:me}
\end{aligned}
\end{equation}
where $\omega_j'=\omega_j + \delta\omega_j$, $\omega_j$ is the transition frequency of atom $j$ only including Lamb shifts from individual connection points, $\delta\omega_j$ is the contribution to the Lamb shift of atom $j$ from interference between connection points, $g$ is the exchange interaction between atoms, $\Gamma_j$ is the individual relaxation rate of atom $j$, and $\Gammacol$ is the collective relaxation rate for the atoms.
A possible external drive is accounted for by $H_d$, the Pauli $Z$ matrix and the raising (lowering) operator of atom $j$ are denoted by $\sigma_z^j$ and $\sigma_+^j (\sigma_-^j)$, respectively, and $\Hc$ denotes Hermitian conjugate.

As shown in \appref{app:me}, the right ($R$) and left ($L$) collapse operators can be written like
\begin{equation}
    \COL_{R/L} = \sqrt{\Gamma_{aR/L}}\,\sma + \sqrt{\Gamma_{bR/L}}\,\smb,
\end{equation}
and they relate to the individual and collective decay rates in the following way:
\begin{align}
    &\Gamma_j=\Gamma_{jR}+\Gamma_{jL},\quad \text{for } j=a,b\\
    &\Gammacol = \sqrt{\Gamma_{aR}^{}\Gamma_{bR}^*} + \sqrt{\Gamma_{aL}^{}\Gamma_{bL}^*}.
\end{align}
Note that the collective decay, $\Gammacol$, is set by interference between emission from connection points belonging to different atoms, whereas the exchange interaction, $g$, is set by emission from connection points of one atom being absorbed at connection points of the other atom~\cite{Kockum18}.

The drive Hamiltonian $H_d$ in \eqref{eq:me} accounts for a coherent drive which, without loss of generality, we assume incoming from the left side and thus propagating towards the right. Then, the drive is quantifiable by a boson flux of $\abs{\beta}^2$ bosons per second, or, equivalently, a Rabi frequency $\Omega_j = 2\beta S_R \sqrt{\Gamma_{jR}^*}$, and the Hamiltonian reads 
\begin{equation}
\begin{aligned}
    H_d =& -i\beta \SCA_R\COL_R^\dagger + \Hc \\
        =&-\frac{i}{2}(\Omega_a\spa + \Omega_b\spb) + \Hc,
    \label{eq:drive}
\end{aligned}
\end{equation}
where $\SCA_R = e^{i\Sigma_k\phi_k}$ is the right-propagating scattering-matrix term and $\Sigma_k\phi_k$ is the sum of the phase shifts acquired between the outermost connection points to the waveguide.
Similarly, we denote the other terms of the system's Hamiltonian $H$ by $H_u$, for undriven Hamiltonian.
In \secsref{sec:dfi}{sec:dsu}, we neglect the drive term (i.e., we set $H_d = 0$ and $H=H_u$); we only consider it in \secref{sec:dsd}.

\input{TableME}

The coefficients $\delta\omega_j$, $\Gamma_j$, $\Gammacol$ and $g$ in \eqref{eq:me} are well-known for small atoms \cite{smallatoms, Lalumiere13, Pichler15} and also for giant atoms in a bidirectional open waveguide~\cite{Kockum18}.
In \tabref{tab:me_ctes}, we provide the coefficients for the chiral case, assuming that the bare relaxation rate at each connection point is $\gamma=\gamma_R+\gamma_L$, where $R$ and $L$ denote right and left, respectively.
From these expressions, a first difference to note with respect to the bidirectional case~\cite{Kockum18} is that $\Gammacol$ and $g$ are no longer always real: they can become complex in the chiral case due to the phase shifts acquired in the waveguide.
In the following sections, we use these expressions to derive conditions where atoms are protected from decoherence, either through decoherence-free interaction (\secref{sec:dfi}) or dark states (\secref{sec:ds}).
Formulas for arbitrary bare relaxation rates and arbitrary phase shifts between connection points are given in \appref{app:me}.

\section{Decoherence-free interaction}
\label{sec:dfi}

One aspect to note from the expressions in \tabref{tab:me_ctes} is that, in general, atoms lose their excitations into the waveguide due to non-zero relaxation rates $\Gamma_j$ and $\Gammacol$.
However, braided giant atoms have the distinctive feature of being able to interact without decohering, i.e., $g\neq 0$ while $\Gamma_j=\Gammacol=0$.
This result was found in Ref.~\cite{Kockum18} using a similar master-equation model, but only for the case of bidirectional coupling. In Ref.~\cite{Carollo2020}, it was shown that this interaction was also possible in waveguides of any chirality, a result that was proved by using a collision model without resorting to a master equation.
Here, we show that we can use the master-equation treatment described in \secref{sec:me} to derive conditions for decoherence-free interaction in waveguides of any chirality.

\begin{figure*}[t]
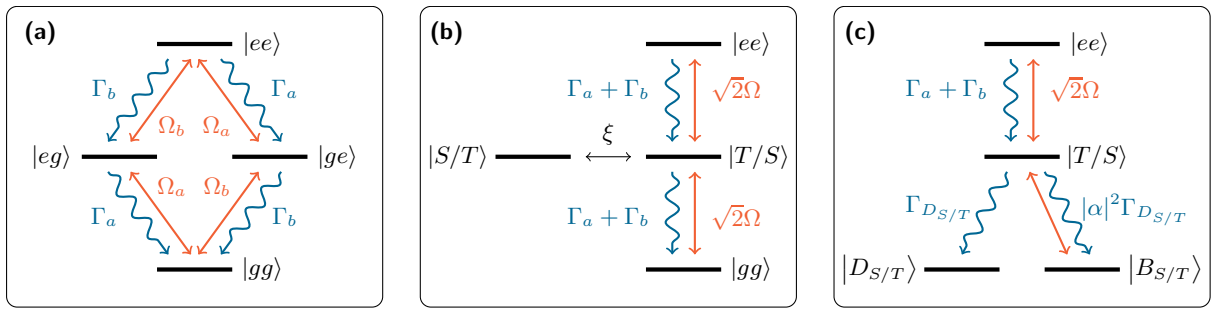

\centering
\includestandalone{Basis}
\caption{Energy-level diagrams of two two-level atoms coupled by a chiral waveguide. (a) The four canonical basis states are that both atoms are excited ($\ket{ee}$), one atom is excited while the other is in its ground state ($\ket{eg}, \ket{ge}$), or both atoms are in their ground state ($\ket{gg}$). The decay rates $\GammaA$ and $\GammaB$, as well as the drives $\OmegaA$ and $\OmegaB$, depend on the topology of the atoms. They are defined in \tabref{tab:me_ctes} and \eqref{eq:drive}, respectively. (b) Equivalent four-state basis, where $\ket{S/T}$ is a dark state which does not decay to the waveguide and $\ket{T/S}$ is a bright state that decays with a superradiant rate $\GammaA+\GammaB$. The states $\ket{S}$ and $\ket{T}$ denote the singlet and the triplet, respectively [see \eqref{eq:singlet_triplet}], and they act as dark or bright states according to the conditions in \tabref{tab:ds}. As derived in \appref{app:ds}, they are coupled by $\xi=(\omegaB'-\omegaA'+g-g^*)/2$, meaning that if $H_d=0$, then $\xi$ must also be zero for $\ket{S/T}$ to be a dark state. If $H_d\neq 0$, then $\xi\neq0$ and we can find a dark state $\ket{D_{S/T}}$ in the subspace spanned by $\ket{gg}$ and $\ket{S/T}$. Also, as shown in \eqref{eq:Hd_eSTg}, the drive couples states $\ket{gg}\leftrightarrow\ket{T/S}\leftrightarrow\ket{ee}$ by $\sqrt{2}(\OmegaA\pm\OmegaB)/2$, which under the dark-state conditions is equal to $\sqrt{2}\OmegaA \equiv \sqrt{2}\Omega$. (c) Diagram for the driven-dissipative regime, where the dark and bright states $\ket{D_{S/T}}$ and $\ket{B_{T/S}}$ are defined in Eqs.~(\ref{eq:d_st}) and (\ref{eq:b_st}), respectively. The rate $\GammaD$ at which they are populated can be found in \eqref{eq:gammaef}. The choice of singlet and triplet states complies with the conditions in \tabref{tab:dsd}, which also contains expressions for $\alpha$.}.
\label{fig:basis}
\end{figure*}

By definition, an atomic ensemble does not decohere into the waveguide when the individual decay rate of each atom is zero, i.e., $\Gamma_j=0\;\forall j$.
In braided atoms, this implies that an excitation acquires a phase $\pi$ ($\mkern-10mu\mod 2\pi$) between the coupling points of atom $j$, or equivalently, that the coupling points are separated by half the excitation's wavelength, $\lambda/2$ ($\mkern-10mu\mod \lambda$).
For this particular distance, the emission from each atom's connection points interferes destructively, making the sum over all atoms zero and thus preventing collective decay ($\Gammacol=0$).
In separate and nested atoms, the connection points of atom $b$ are consecutive, so the emission between them cancels ($g=0$) when $\GammaB=0$.
Unlike these topologies, braided atoms have the particularity that no consecutive points belong to the same atom, allowing a non-zero exchange interaction.
This implies that an excitation can be released from atom $a$ to be reabsorbed by atom $b$ and vice versa, in a continuous loop.

We observe that this reasoning applies separately in each propagation direction and thus it holds for waveguides of any chirality. Mathematically, the decoherence-free interaction conditions for two braided giant atoms read
\begin{equation}
    \mqty{\phi_1+\phi_2 = (2n_1+1)\pi,\\ \phi_2+\phi_3=(2n_2+1)\pi,}\quad n_1,n_2\in\mathbb{N},
\end{equation}
which yield $g=(2\gamma_L e^{i\phi_2}-2\gamma_R e^{-i\phi_2})/(2i)$, according to \tabref{tab:me_ctes}. Therefore, for bidirectional coupling, the additional condition $\phi_2\neq n_3\pi$ with $n_3\in\mathbb{N}$ is required in order to obtain nonzero interaction.

The reasoning is also valid for arbitrary bare relaxation rates, as shown in Ref.~\cite{Kockum18} for bidirectional coupling. Nevertheless, in such a case, $\Gamma_j$ is only zero for both atoms when $\gamma_{1R/L}=\gamma_{3R/L}$ and $\gamma_{2R/L}=\gamma_{4R/L}$.

\section{Dark states}
\label{sec:ds}

\subsection{Undriven system}
\label{sec:dsu}

An important aspect of the decoherence-free interaction in \secref{sec:dfi} is that it is independent of the states of the giant atoms, meaning that the entire Hilbert space of the atomic ensemble is protected from decoherence.
A less robust way of preventing atoms from decaying is through dark states, whereby only a subspace of the Hilbert space is protected from decoherence.
In an atomic network, dark states are nonradiative pure states which are annihilated by all collapse operators and are eigenstates of the multiatom Hamiltonian~\cite{Lenz93, Lalumiere13,  Stannigel12, Pichler15, Guimond16}.

In the case of two small atoms coupled nonchirally, a non-trivial dark state (i.e., not the ground state) arises when the atoms share an excitation that is prevented from decaying by destructive interference between coupling points.
This behaviour, however, is not possible if the atoms are coupled to a unidirectional waveguide, wherein only one of the atoms is ``aware" of the other~\cite{Lodahl17}.
Therefore, in general, when the system is not driven [$H_d=0$ in \eqref{eq:me}], chirally coupled small atoms do not exhibit dark states.
In this section, we look for dark states in setups with two giant atoms and find that, in the nested configuration, they can share an excitation even when they are chirally coupled.

Let us consider two qubits $a$ and $b$ with ground states $\ket{g}$ and excited states $\ket{e}$.
The canonical states of the ensemble can then be described by the basis $\{\ket{gg},\, \ket{ge},\, \ket{eg},\, \ket{ee}\}$, where $\ket{nm}=\ket{n}_a\otimes\ket{m}_b$ is the tensor product of the state of both atoms [see \figpanel{fig:basis}{a}].
A dark state $\ket{D}$ is therefore, in general, a linear combination of the four states above. It is derived from the aforementioned definition, which formally reads
\begin{equation}
\begin{aligned}
    &\COL_R\ket{D}=\COL_L\ket{D}=0\\
    &H\ket{D} = \mu\ket{D},\;\, \mu\in\R.
    \label{eq:ds_conditions}
\end{aligned}
\end{equation}
In particular, the first condition ensures that $\ket{D}$ is nonradiative and the second one ensures it is a steady state.

Taking the undriven Hamiltonian of the system $H=H_u$ and the collapse operators $\COL_{R/L}$ defined in \secref{sec:me}, we find that under certain restrictions, the dark state can either be the singlet state $\ket{S}$ or the triplet $\ket{T}$:
\begin{equation}
\begin{aligned}
     \ket{S} = \frac{1}{\sqrt{2}}\mleft(\ket{ge}-\ket{eg}\mright)\\
     \ket{T} = \frac{1}{\sqrt{2}}\mleft(\ket{ge}+\ket{eg}\mright).
     \label{eq:singlet_triplet}
\end{aligned}
\end{equation}
These restrictions are collected in \tabref{tab:ds}, where we show that the dark state is $\ket{S}$ or $\ket{T}$ depending on the phase acquired by the traveling boson in the waveguide.
Note that, for simplicity, we have derived the constraints assuming the bare relaxation rates are equal at all connection points, but different in each propagation direction.
In all atomic setups, this perfect subradiance can only be achieved if both atoms have the same transition frequency ($\omegaA=\omegaB$) and, with the exception of nested atoms, only if the waveguide is bidirectional ($\gamma_R=\gamma_L$).
A key result that we find is that nested atoms allow for the existence of a dark state regardless of the chirality of their coupling to the waveguide.

\input{TableDS}

Taking a closer look at the dark-state conditions in \tabref{tab:ds}, we see that nested atoms require $\phi_2\neq\pi$, i.e., they require atom $b$ to be able to decay ($\GammaB\neq 0$).
If this atom does not relax, it is also not able to absorb an excitation, and $\Gammacol, g=0$.
Then, if $\phi_1$ and $\phi_3$ are chosen such that atom $a$ cannot decay either, the diatomic ensemble does not decohere, but it also does not interact.
In such a situation, all the states are dark and the system does not evolve. Therefore, setting $\phi_2\neq\pi$ allows atom $b$ to release an excitation which can be absorbed by atom $a$, even in chiral waveguides.
If we then choose $\phi_1=\phi_3$, this release is compensated by the possibility of atom $a$ relaxing to excite atom $b$, thus making the maximum entanglement between $\ket{ge}$ and $\ket{eg}$ a dark state.
Note that the reason why the other topologies do not have a dark state in chiral waveguides is because atom $b$ cannot excite atom $a$ with the same probability as $a$ can excite $b$. In fact, this distinctive symmetry that allows nested atoms to have a dark state regardless of the chirality is also present when a giant atom $a$ nests a small atom $b$ between its connection points.

Another significant difference between small and giant atoms exists with regard to trivial dark states.
In small atoms, only $\ket{gg}$ is a trivial dark state, whereas in giant atoms, we can also find phase shifts that make all states dark.
In this case, the atoms are completely decoupled from the waveguide and each other.
This regime should be accessible in architectures with tunable atomic frequencies (which changes the phases $\phi_j$)~\cite{Kockum14, Kannan20}, such as superconducting qubits coupled to a transmission line.

As derived in \appref{app:ds} and depicted in \figpanel{fig:basis}{b}, an alternative multiatom state basis is $\{\ket{gg},\, \ket{S},\, \ket{T},\, \ket{ee}\}$, where $\ket{S}$ and $\ket{T}$ are coupled by $\xi=(\omegaB'-\omegaA'+g-g^*)/2$. Whenever $\xi=0$, there can be a dark state $\ket{D} = \ket{S/T}$ that does not relax into the waveguide, and a bright state $\ket{B} = \ket{T/S}$ that decays at a superradiant rate $\GammaA+\GammaB$.
We use the notation $\ket{S/T},\,\ket{T/S}$ to denote that when $\ket{D}=\ket{S}$ in the instances presented in \tabref{tab:ds}, then $\ket{B}=\ket{T}$ and vice versa. Let us remark that $\xi=0$ when $g\in\mathbb{R}$ and $\omegaA'=\omegaB'$, which ultimately imposes the chirality and frequency restrictions from \tabref{tab:ds}.

\input{TableDSd}

\input{TableGamma}

\subsection{Driven-dissipative regime}
\label{sec:dsd}

In Ref.~\cite{Pichler15}, it is shown that one can get around the absence of dark states for small atoms in chiral waveguides by coherently driving the system.
When such a drive is added, the diatomic ensemble evolves to a dynamic equilibrium between drive and dissipation where the stream of bosons scattered from the first atom interferes destructively with the bosons scattered from the second~\cite{Lodahl17}.
Here, we show that such a regime is also accessible for all three topologies of giant atoms.

We consider an incoming coherent signal from the left side of the waveguide, as described in \eqref{eq:drive}. Then, with the modified Hamiltonian $H=H_u+H_d$, it follows from the conditions in \eqref{eq:ds_conditions} that the dark state in the driven-dissipative regime is one of the following:
\begin{equation}
    \ket{D_{S/T}} = \frac{1}{\sqrt{1 + \abs{\alpha}^2}} \mleft(\alpha\ket{S/T} + \ket{gg}\mright),
    \label{eq:d_st}
\end{equation}
where $\alpha= i\sqrt{2}\Omega/ (2\xi)$ is a function of the bare relaxation rates, the coherent drive, and the detuning of the atoms from the drive (see \tabref{tab:dsd}).
These dark states exist in all topologies regardless of the chirality of the waveguide the atoms are coupled by, but only under the constraints collected in \tabref{tab:dsd}.
In particular, if the atoms are detuned from the drive frequency by $\delta_j$, then $\ket{D_{S/T}}$ only exists if the detunings are opposite (i.e., $\deltaA = -\deltaB$), and additionally, the detuning must be nonzero for nested atoms.
Moreover, from the phase shifts in \tabref{tab:dsd} it follows that the drive amplitudes of both atoms are equal ($\OmegaA=\OmegaB$) when $\ket{D_S}$ is dark and opposite ($\OmegaA=-\OmegaB$) when $\ket{D_T}$ is dark. We use the notation $\Omega\equiv\OmegaA$ and $\delta\equiv\deltaA$ to refer to the drive and detuning, respectively. 

Note that the bright states accompanying the dark states above are
\begin{equation}
    \ket{B_{S/T}} = \frac{1}{\sqrt{1 + \abs{\alpha}^2}} \mleft(\ket{S/T}-\alpha^*\ket{gg}\mright).
    \label{eq:b_st}
\end{equation}
The existence of these dark and bright states $\ket{D_{S/T}}$ and $\ket{B_{S/T}}$ is intuitive from the fact that the diagram of states $\ket{gg}$, $\ket{S/T}$ and $\ket{T/S}$ in \figpanel{fig:basis}{b} mimics a $\Lambda$ system \cite{Ask20, Pichler15, Guimond16}. Thus, when the atoms are externally driven, a new dark state arises in the subspace spanned by $\ket{S/T}$ and $\ket{gg}$.

In the driven-dissipative regime, it is more convenient to write the master equation in the basis $\{\ket{D_{S/T}},\, \ket{B_{S/T}},\, \ket{T/S},\, \ket{ee}\}$, as depicted in \figpanel{fig:basis}{c}.
In \appref{app:dsd_driven}, we show that this change of basis allows us to derive the rate $\GammaD$ at which the dark state is populated:
\begin{equation}
    \GammaD = \frac{\GammaA+\GammaB}{1+\abs{\alpha}^2}.
    \label{eq:gammaef}
\end{equation}
In particular, we write the full expression of $\GammaD$ for each setup in \tabref{tab:gamma}, from which we observe that giant atoms can populate the driven-dissipative dark state faster than small atoms.

It is not trivial to quantify how much faster the dark state can be populated in a giant atom than in a small one. 
By keeping the same drive amplitude $\Omega$, we find that $\eval{\Gamma_{D_{S/T}}(\Omega)}_\text{giant} \le 64\eval{\Gamma_{D_{S/T}}(\Omega)}_\text{small}$. 
However, that might seem like an unfair comparison, since $\Omega$ contains interference effects between the connection points of the same atom which are not present in small emitters.
A fairer comparison might be made by assuming the same boson flux $\abs{\beta}^2$.
By looking at the expressions in \tabref{tab:gamma}, we see that in the limiting case where the drive $\gamma_R\abs{\beta}^2$ is weak compared to the detuning $\delta^2$ and the coupling-strength asymmetry $\Delta\gamma^2$, then  $\eval{\Gamma_{D_{S/T}}(\beta; \gamma)}_\text{giant} \le 4\eval{\Gamma_{D_{S/T}}(\beta; \gamma)}_\text{small}$.
However, by evaluating the rates numerically, it can be seen that when the drive strength is comparable to the coupling-strength asymmetry, $\eval{\Gamma_{D_{S/T}}(\beta; \gamma)}_\text{separate}$ can be up to $16\eval{\Gamma_{D_{S/T}}(\beta; \gamma)}_\text{small}$.
Nevertheless, this might again be an unfair comparison since separate atoms have twice as many connection points as small atoms with the same decay rate.
If instead we consider the small atoms to decay with rate $4\gamma$ (not $2\gamma$, due to the square root in the jump operator), then we again recover $\eval{\Gamma_{D_{S/T}}(\beta; \gamma)}_\text{giant} \le 4\eval{\Gamma_{D_{S/T}}(\beta; \gamma)}_\text{small}$.

Hence, we conclude that regardless of the choice of comparison, giant atoms can populate the driven-dissipative dark state faster than small atoms. This feature might be advantageous in typically slow processes such as adiabatic-passage techniques \cite{Bergmann98, Bergmann15, Vitanov17}.

\section{Multiple atoms with multiple coupling points}
\label{sec:general}

Now that we have illustrated the main mechanisms to avoid decoherence with two small and giant atoms, it is natural to generalize the dynamics to multiple atoms with multiple connection points each. In the same way we have built a model in \secref{sec:me} for two atoms, we can derive a master equation for an arbitrary number of atoms $N$, where each atom $j$ has $M_j$ connection points.
In this case, following the procedure described in \appref{app:me}, we obtain the master equation
\begin{widetext}
\begin{equation}
\begin{aligned}
    \dot{\rho} =&
	-i \comm{ \sum_{j=1}^N \omega_j'  \frac{\sigma_z^j}{2} + \sum_{j=1}^{N-1}\sum_{k=j+1}^N (g_{j,k}\sigma_-^j\sigma_+^k + \Hc)
	+ H_d}{\rho} + \sum_{j=1}^N \Gamma_j \D[\sigma_-^j]\rho \\
	& +\sum_{j=1}^{N-1}\sum_{k=j+1}^N \mleft[\Gamma_{\text{coll},j,k} \mleft(\sigma_-^j\rho\sigma_+^k-\frac{1}{2}\acomm{\sigma_-^j\sigma_+^k}{\rho}\mright) + \Hc \mright],
	\label{eq:me_multi}
\end{aligned}
\end{equation}
with the constants
\begin{align}
	\delta\omega_j =& \sum_{n=1}^{M_j-1}\sum_{m=n+1}^{M_j} \mleft(\sqrt{\gamma_{j_nR}\gamma_{j_mR}} +\sqrt{\gamma_{j_nL}\gamma_{j_mL}}\mright) \sin(\phi_{j_n,j_m}),\\
	g_{j,k} =&  \sum_{n=1}^{M_j} \sum_{m=1}^{M_k}  \frac{\varepsilon}{2i} \mleft[ \sqrt{\gamma_{j_nR}\gamma_{k_mR}}\, \exp(\varepsilon i \phi_{j_n,k_m}) - \sqrt{\gamma_{j_nL}\gamma_{k_mL}}\, \exp(- \varepsilon i \phi_{j_n,k_m}) \mright],\\
	\Gamma_j =& \sum_{n=1}^{M_j} \sum_{m=1}^{M_j} \mleft( \sqrt{\gamma_{j_nR}\gamma_{j_mR}} + \sqrt{\gamma_{j_nL}\gamma_{j_mL}} \mright) \cos(\phi_{j_n,j_m}),\\
	\Gamma_{\text{coll}, j,k} =& \sum_{n=1}^{M_j} \sum_{m=1}^{M_k}  \mleft[ \sqrt{\gamma_{j_nR}\gamma_{k_mR}}\, \exp(\varepsilon i \phi_{j_n,k_m}) + \sqrt{\gamma_{j_nL}\gamma_{k_mL}}\, \exp(- \varepsilon i \phi_{j_n,k_m}) \mright],
	\label{eq:me_ctes_multi}
\end{align}
\end{widetext}
where $\phi_{m,n} = \sum_{k=m}^n \phi_k$,  the position of connection point $j_n$ is denoted by $x_{j_n}$, and 
\begin{equation}
    \varepsilon = \begin{cases} +1 & \text{ if } x_{j_n}<x_{k_m}\\
    \;\;\,0 & \text{ if } x_{j_n}=x_{k_m}\\
    -1 & \text{ if } x_{j_n}>x_{k_m}. \end{cases}
    \label{eq:epsilon}
\end{equation}
In particular, for $N=2$ and $M_{1,2} \le 2$ we retrieve \eqref{eq:me} with the coefficients from \tabref{tab:me_ctes_full}.

When there are multiple atoms with multiple connection points, we can still distinguish the three giant-atom layouts displayed in \figref{fig:setups}.
By taking the atoms in pairs, we define the \emph{separate} configuration as the one where all connection points of atom $j$ are situated to the left of all connection points of atom $k$, and the \emph{nested} configuration as the one where all coupling points of atom $k$ are situated between two consecutive points of atom $j$.
The \emph{braided} topology encompasses all other configurations, i.e., the cases where some connection points of atom $j$ are between some connection points of atom $k$.

\subsection{Decoherence-free interaction}

A first thing to notice from \eqref{eq:me_multi} is that all interactions are pairwise, which means that the phenomena derived for two atoms in the preceding sections can be generalized to multiple atoms. For instance, in order to achieve decoherence-free interaction between multiple atoms, it is necessary to have $g_{j,k}\neq0$ and $\Gamma_j, \Gamma_{\text{coll},j,k}=0$ for all $j,k$, which by the definition above is still only possible in braided configurations. Therefore, interesting layouts such as a 1D spin chain with protected nearest-neighbour interaction [\figpanel{fig:dfi_multi}{a}] or an atomic triad with protected all-to-all connectivity [\figpanel{fig:dfi_multi}{b}]~\cite{Kockum18} can be built with waveguides of any chirality.

Let us remark that decoherence-free interaction is also achievable with more than two coupling points, as shown in Ref.~\cite{Carollo2020}.
For instance, the three-coupling-point architecture used in Ref.~\cite{Kannan20} to demonstrate decoherence-free interaction in a bidirectional waveguide is also protected from decay in a chiral waveguide.

\subsection{Dark states}

In \secref{sec:dsu}, we showed that two atoms with two connection points each can share an excitation and form a dark state, but only nested atoms can do so with chiral coupling due to their distinctive symmetry. This property can be generalized to multiple atoms with multiple coupling points through the following sufficient conditions for the existence of a dark state:
\begin{enumerate}
    \item $\sqrt{\Gamma_{jR}}/\sqrt{\Gamma_{kR}} = \sqrt{\Gamma_{jL}}/\sqrt{\Gamma_{kL}} \;\forall j,k$ and for certain phase shifts. This ensures that a dark state $\ket{D}$ exists such that $\COL_R\ket{D}=\COL_L\ket{D}=0$ for those phase shifts. Note that if $\sqrt{\Gamma_{jR/L}}=0 \;\forall j$, then all states are dark.
    \item $H\ket{D}=\mu\ket{D},\; \mu\in\mathbb{R}$. By definition, a dark state is an eigenstate of the Hamiltonian.
    \item $\omega_j'=\omega_k'\;\forall j,k$. One way to ensure this, but not necessary, is setting $\omega_j=\omega_k$ and $\delta_j=0\;\forall j,k$.
    \item $g_{j,k}\in\mathbb{R}\; \forall j,k$. This condition, together with the previous one, ensures that the dark and bright states are decoupled, as explained in \secref{sec:dsu}, where we set $\xi=0$.
\end{enumerate}

Some symmetric layouts that obey these conditions are ``matryoshka'' nested atoms [\figpanel{fig:ds_multi}{a}], which, like the eponymous Russian doll, are iteratively nested one inside another, or enclosed braided atoms [\figpanel{fig:ds_multi}{b}], which are braided, but have the two outermost connection points belonging to the same atom.
For instance, if we set all the phases to $0 \mkern-5mu\mod 2\pi$ and all the relaxation rates to $\gamma = \gamma_R + \gamma_L$, then three matryoshka atoms can share one excitation in the state $\ket{D}=(\ket{egg}+\ket{geg}-2\ket{gge})/\sqrt{6}$, and two enclosed braided atoms can share an excitation in the state $\ket{D} = (2\ket{eg}-3\ket{ge})/\sqrt{13}$.

\begin{figure}[t]
\centering
\includestandalone{Chain}
\includestandalone{Triad}
\caption{Examples of atomic setups that allow decoherence-free interaction ~\cite{Kockum18} even when coupled to a chiral waveguide. (a) 1D braided chain with protected nearest-neighbour interaction. (b) Braided triad with protected all-to-all connectivity.\\}
\label{fig:dfi_multi}

\includestandalone{Matryoshka}
\includestandalone{Enclosed}
\caption{Examples of undriven atomic setups that have dark states even when coupled to a chiral waveguide. (a) Matryoshka nested atoms, which illustrate that the number of atoms does not affect the existence of a dark state, only its form. (b) Enclosed braided atoms, which show that atoms with different numbers of connection points can also be entangled and share an excitation to form a dark state.\\}
\label{fig:ds_multi}

\includestandalone{Splitting1}
\includestandalone{Splitting2}
\caption{Example of the invariance of a dark state under splitting of connection points. (a) Braided configuration which exhibits a dark state when $\phi=0, \pi$. (b) Splitting of the last connection point in layout (a) into $n$ consecutive connection points with coupling strengths $\gamma_1,\dots,\gamma_n$. This layout exhibits a dark state provided that $\mleft(\sum_{j=1}^n\sqrt{\gamma_j}\mright)^2=\gamma$ and no additional phase shift \mbox{($\mkern-10mu\mod 2\pi$)} is acquired in the splitting.}
\label{fig:splitting}
\end{figure}

The existence of a dark state is invariant under the splitting of connection points, provided that if one point with relaxation rate $\gamma$ splits into $n$ consecutive points with coupling strengths $\gamma_1,\dots, \gamma_n$, then $\mleft(\sum_{j=1}^n\sqrt{\gamma_j}\mright)^2=\gamma$ and there is no additional phase shift between points ($\mkern-10mu\mod 2\pi$).
Such a case is illustrated in \figref{fig:splitting}, where \figpanel{fig:splitting}{a} and \figpanel{fig:splitting}{b} depict a braided layout before and after splitting, respectively.

 \section{Conclusion and outlook}
\label{sec:conclusion}

In this paper, we have derived a master equation for multiple giant atoms chirally coupled to a 1D open waveguide at multiple points.
We have shown that, when giant atoms have their connection points in a braided configuration, they can interact without decohering into the waveguide. 
This exchange interaction is mediated by the waveguide and is not protected from decay in small atoms (nor in separate and nested giant atoms). In braided giant atoms, however, the protection is possible regardless of the chirality of the waveguide and the atomic state of the system.

Moreover, we have shown that giant atoms exhibit dark states in both the static and driven-dissipative regimes, and we have derived conditions for the existence of such subradiant states.
In particular, we have found that, unlike small atoms, undriven giant atoms in the nested and braided configurations can present dark states regardless of the chirality of the waveguide. We have also shown that dark states of coherently driven giant atoms can be populated faster than their counterparts in small atoms.

This work has potential applications in quantum simulations and quantum networks in chiral settings, as discussed and demonstrated in the experiments of Refs.~\cite{Kannan20, Vadiraj21} for nonchiral waveguides. As shown in Ref.~\cite{Kannan20}, the decoherence-free interaction enables the creation of any quantum many-body state among the atoms, since the interaction can be used to form a universal set of quantum gates. The interaction of this many-body system with the environment represented by the waveguide can then be turned on and off for quantum simulations by controlling the phase shifts $\phi_j$ in the setup, e.g., by tuning the atomic frequencies. Similarly, the universal gate set can be used to create entangled states that then are released into the waveguide for quantum communication or one-way quantum computing. Furthermore, the dark states we have found here for giant atoms could find applications in quantum memories and quantum networks, as described in the introduction.

Future work may include studying the driven-dissipative regime beyond the diatomic case presented here, to observe the effect of chirality on the formation of dimers, tetrameres, and other multipartite entangled states~\cite{Pichler15, Guimond16}.
The theory derived here may also be extended by considering time delay between connection points~\cite{Guo17, Andersson2019, Guo20, Guo2020a, Du2021}, e.g., to determine how robust the decoherence-free interaction or the dark states are to such time delays.
Another interesting path to explore might be to make the atoms three-level $\Xi$, $\Lambda$, $V$, or $\Delta$ systems~\cite{Scheucher20, Guimond16}.
Finally, we believe that the phenomena found here unique to giant atoms in a chiral setting are ready to be demonstrated experimentally in several experimental platforms, e.g., using superconducting qubits coupled to a transmission line with circulators inserted to provide the chirality.

\begin{acknowledgments}
We acknowledge support from the Swedish Research Council (grant number 2019-03696) and from the Knut and Alice Wallenberg Foundation through the Wallenberg Centre for Quantum Technology (WACQT).
\end{acknowledgments}

\appendix
\section{Derivation of the master equation}
\label{app:me}

\input{TableL}

\subsection{Two atoms with one or two connection points}

In the model presented in \secref{sec:me}, we use the Lindblad form of the master equation \cite{OQS_book, Carmichael_book}.
For this to be valid, we need to make several assumptions.
First, we make the Born approximation, which relies on the coupling between the atom and the waveguide being weak. This implies that the density matrix of the waveguide does not change significantly due to the interaction with the atoms.
Secondly, we make the Markov approximation, through which we assume that the waveguide has no memory, i.e., that any effect that an atom has on the waveguide at a certain time does not affect the dynamics of the atom at a later time. This relies on the assumption that the travel time between the connection points of the giant atoms is negligible compared to the relaxation times of the atoms.
With these two approximations, we can further apply the rotating-wave approximation (RWA) and trace over the waveguide modes to obtain the master equation in the Lindblad form, as written in \eqref{eq:me_Lindblad} in the main text:
\begin{equation}
    \dot{\rho} = -i[H, \rho] + \sum_{k=1}^n   \D[\COL_k]\rho,
    \label{eq:me_Lindblad_app}
\end{equation}
where $\D[X]\rho = X\rho X^\dag-\frac{1}{2}\acomm{X^\dag X}{\rho}$.

In order to find the Hamiltonian $H$ and collapse operators $\COL_k$ of the atomic ensemble, we use the SLH formalism for cascaded systems \cite{SLH1, SLH2, SLH3}. In the way we present it here, this formalism requires the Born and Markov approximations, that the bosons propagate in a linear medium without dispersion, and that the travel time between connection points is negligible compared to the relaxation times of the atoms.

In the SLH formalism, an open quantum system with $n$ input-output ports is described by a triplet $G=(\SCA,\, \COL,\, H)$, where $\SCA$ is an $n\times n$ scattering matrix, $\COL$ is an $n\times 1$ vector representing the coupling between the system and the environment at the input-output ports, and $H$ is the Hamiltonian of the system.
In our case, $\COL$ models the coupling between an atom $j$ and the waveguide at a connection point $k$, in such a way that $\COL_k=\sqrt{\gamma_k}\sigma_-^j$.

We proceed in the following way, similar to the derivation for giant atoms in a bidirectional waveguide in the supplementary material of Ref.~\cite{Kockum18}.
At each connection point $k$, we define an SLH triplet:
\begin{equation}
    G_k =\begin{cases}
    \mleft( 1,\, \sqrt{\gamma_k}\sigma_-^j,\, \frac{1}{2}\omega_j\sigma_z^j\mright) & \text{if } k \text{ is the first coupling}\\[-4pt]
    &\text{point of atom } j\\[4pt]
    \mleft(1,\, \sqrt{\gamma_k}\sigma_-^j,\, 0\mright) & \text{otherwise.}
    \end{cases}
\end{equation}
To account for the phase shift acquired between connection points $k$ and $k+1$, we define $G_{\phi_k}=(e^{i\phi_k},\, 0,\, 0)$.
We then take one propagation direction and apply a \emph{series product} between all the triplets, as if the system was cascaded.
In the SLH formalism, the series product between two triplets is defined as follows:
\begin{align}
    G_2\series G_1 = \Big(&\SCA_2\SCA_1,\;\; \SCA_2\COL_1 + \COL_2,\\ &H_1+H_2+\frac{1}{2i}\mleft(\COL_2^\dagger \SCA_2\COL_1-\COL_1^\dagger \SCA_2^\dagger \COL_2\mright)\Big).\nn
    \label{eq:series_product}
\end{align}

If the waveguide is unidirectional, the series product yields the final triplet.
Otherwise, we repeat the process in the other propagation direction, which results in two triplets, $G_R$ and $G_L$.
Since propagation to the right and left directions occurs simultaneously, we can \emph{concatenate} the two triplets according to SLH practice:
\begin{equation}
    G_1\boxplus G_2 = \mleft( \mqty(\SCA_1 & 0\\0 & \SCA_2), \mqty(\COL_1\\\COL_2), H_1 + H_2\mright).
    \label{eq:concatenation}
\end{equation}

Therefore, for all the setups depicted in \figref{fig:setups}, the final SLH triplet has the following form.
The scattering matrix reads
\begin{equation}
    \SCA = \mqty(\SCA_R & 0 \\ 0 & \SCA_L) = \mqty(e^{i\Sigma_k\phi_k} & 0 \\ 0 & e^{i\Sigma_k\phi_k}),
    \label{eq:S}
\end{equation}
where $\Sigma_k\phi_k$ is the sum of all the phase shifts acquired in the waveguide. The collapse operator, also known as coupling vector or jump operator, reads
\begin{equation}
    \COL = \mqty(\COL_R + \COL_d\\\COL_L),
    \label{eq:L}
\end{equation}
where $\COL_{R/L} = \sqrt{\Gamma_{aR/L}}\,\sma + \sqrt{\Gamma_{bR/L}}\,\smb$ and $\COL_d=0$ if there is no external drive. In particular, we write $\COL_{R/L}$ for each topology as shown in \tabref{tab:L}.

Finally, the last component of the SLH triplet is the Hamiltonian of the system, which reads
\begin{equation}
    H = \frac{1}{2}\omegaA'\sza + \frac{1}{2}\omegaB'\szb + g \smapb + g^* \spamb + H_d,
    \label{eq:H}
\end{equation}
where $\omega_j'$ and $g$ obey the expressions in \tabref{tab:me_ctes_full} and $g^*$ denotes the complex conjugate of the exchange interaction $g$.
Using the series product and concatenation from the SLH formalism, we find that, given an incoming coherent drive from the left, with a boson flux of $\abs{\beta}^2$, then
\begin{equation}
\begin{aligned}
    \COL_d &= \beta \SCA_R\\
    H_d &=\frac{-i}{2}\mleft(\beta \SCA_R\COL_R^\dagger - \beta^*\SCA_R^\dagger \COL_R\mright).
    \label{eq:drive1}
\end{aligned}
\end{equation}

\input{TableME_full}

Using \eqsref{eq:L}{eq:drive1}, we can expand the master equation in the Lindblad form
[\eqref{eq:me_Lindblad_app}] to obtain
\begin{equation}
\begin{aligned}
	\dot{\rho} = & -i\comm{H}{\rho} + \D[\COL_R+\COL_d]\rho + \D[\COL_L]\rho =\\
	=& -i \comm{\omegaA'\frac{\sza}{2} + \omegaB'\frac{\szb}{2} + \mleft(g\smapb +
	\Hc\mright) + H_d}{\rho}\\
	&+ \GammaA \D[\sma]\rho + \GammaB \D[\smb]\rho \\
	& +\mleft[\Gammacol \mleft(\sma\rho\spb-\frac{1}{2}\acomm{\smapb}{\rho}\mright) + \Hc \mright],
	\label{eq:me_app}
\end{aligned}
\end{equation}
which we note is the same as \eqref{eq:me}, with the constants from \tabref{tab:me_ctes_full}.
It can be shown that the dynamics of a system with this master equation, and the collapse operator and Hamiltonian from \eqref{eq:drive1}, are the same that a system whose drive operators are
\begin{equation}
\begin{aligned}
    \COL_d & = 0\\ 
    H_d & = -i\mleft(\beta \SCA_R\COL_R^\dagger - \beta^*\SCA_R^\dagger \COL_R\mright).
    \label{eq:drive2}
\end{aligned}
\end{equation}
For convenience, we use the latter in the derivation of the dark states in the driven-dissipative regime.

\subsection{Multiple atoms with multiple coupling points}

We can use the same procedure described above to derive the SLH triplet of an arbitrary number of atoms $N$, where each atom $j$ has $M_j$ connection points. The result is
\begin{align}
    \label{eq:S_multi}
    \SCA =& \mqty(\exp(i\phi_{1_1,N_{M_N}}) & 0 \\ 0 & \exp(i\phi_{1_1,N_{M_N}}))\\
    \COL =& \mqty(\displaystyle
    \sum_{j=1}^N \sum_{n=1}^{M_j} \exp(i \phi_{j_n, N_{M_N}}) \sqrt{\gamma_{j_nR}} \,\sigma_-^j\\ \displaystyle
    \sum_{j=1}^N \sum_{n=1}^{M_j} \exp(i \phi_{1_1, j_n}) \sqrt{\gamma_{j_nL}} \,\sigma_-^j)
    \label{eq:L_multi}
\end{align}

\begin{widetext}
\begin{align}
    H =& \sum_{j=1}^N \left[ \omega_j + \underbrace{\sum_{n=1}^{M_j-1}\sum_{m=n+1}^{M_j} \left(\sqrt{\gamma_{j_nR}\gamma_{j_mR}} +\sqrt{\gamma_{j_nL}\gamma_{j_mL}}\right) \sin(\phi_{j_n,j_m})}_{\delta\omega_j} \right] \frac{\sigma_z^j}{2}\nn\\
    &+ \sum_{j=1}^{N-1} \sum_{k=j+1}^N \left(\sigma_-^j\sigma_+^k \underbrace{ \left\{  \sum_{n=1}^{M_j} \sum_{m=1}^{M_k}  \frac{\varepsilon}{2i} \left[ \sqrt{\gamma_{j_nR}\gamma_{k_mR}}\, \exp(\varepsilon i \phi_{j_n,k_m}) - \sqrt{\gamma_{j_nL}\gamma_{k_mL}}\, \exp(- \varepsilon i \phi_{j_n,k_m}) \right] \right\}}_{g_{j,k}} + \Hc \right) ,
    \label{eq:H_multi}
\end{align}
\end{widetext}
where $\phi_{m,n} = \sum_{k=m}^n \phi_k$ and 
\begin{equation}
    \varepsilon = \begin{cases} +1 & \text{ if } x_{j_n}<x_{k_m}\\
    \;\;\,0 & \text{ if } x_{j_n}=x_{k_m}\\
    -1 & \text{ if } x_{j_n}>x_{k_m}. \end{cases}
    \label{eq:app_epsilon}
\end{equation}
This SLH triplet yields the Lindblad master equation shown in the main text [\eqref{eq:me_multi}].

\section{Hamiltonian in the dark-state basis}
\label{app:ds}

\subsection{Undriven system}
\label{app:ds_undriven}

When the diatomic system is not driven, the Hamiltonian can be written in the basis $\{\ket{gg}, \ket{S}, \ket{T}, \ket{ee}\}$ as follows:
\begin{equation}
\begin{aligned}
    H =& H_u = \frac{1}{2}\omegaA'\sza + \frac{1}{2}\omegaB'\szb + g \smapb + g^* \spamb =\\
      =& \frac{\mleft(\omegaA'+\omegaB'\mright)}{2}\ketbra{ee}- \frac{\mleft(\omegaA'+\omegaB'\mright)}{2}\ketbra{gg} \\
      &+\frac{(g+g^*)}{2}\ketbra{T}{T} - \frac{(g+g^*)}{2}\ketbra{S}{S}\\
      &+\frac{(\omegaB'-\omegaA'+g-g^*)}{2}\ketbra{S}{T} +\Hc,
      \label{eq:H_ST_app}
\end{aligned}
\end{equation}
taking $\ket{S}$ and $\ket{T}$ from \eqref{eq:singlet_triplet} in the main text, using the definition $\sigma_z^j = \ket{e}_j\bra{e}_j-\ket{g}_j\bra{g}_j$, and allowing the abuse of notation $\sza \equiv \sza\otimes\mathbb{I}_2$, $\szb \equiv \mathbb{I}_2\otimes\szb$.
 With the Hamiltonian in this form, it is straightforward to see from the last two terms that, whenever $\omegaA'\neq\omegaB'$ or $g\notin\R$, the singlet and triplet states are coupled and none of them can be dark states.
 
This can also be mathematically derived using the dark-state conditions from \eqref{eq:ds_conditions}:
\begin{equation}
\begin{aligned}
    & \COL_R\ket{D}=\COL_L\ket{D}=0 \\
    & H\ket{D} = \mu\ket{D},\;\, \mu\in\R.
    \label{eq:ds_conditions_app}
\end{aligned}
\end{equation}
 Since
\begin{equation}
\begin{aligned}
    \COL_{R/L} = &\sqrt{\Gamma_{aR/L}}\,\sma + \sqrt{\Gamma_{bR/L}}\,\smb= \\
    = & \frac{\sqrt{2}}{2}\mleft(\sqrt{\Gamma_{aR/L}}-\sqrt{\Gamma_{bR/L}}\,\mright)
    \mleft(\ketbra{S}{ee}-\ketbra{gg}{S}\mright) \\
    & +\frac{\sqrt{2}}{2}\mleft(\sqrt{\Gamma_{aR/L}}+\sqrt{\Gamma_{bR/L}}\,\mright) \mleft(\ketbra{T}{ee}+\ketbra{gg}{T}\mright),
    \label{eq:L_ST_app}
\end{aligned}
\end{equation}
the first condition implies that $\ket{D}=\ket{S}$ when $\sqrt{\Gamma_{aR/L}}=\sqrt{\Gamma_{bR/L}}$ and $\ket{D}=\ket{T}$ when $\sqrt{\Gamma_{aR/L}}=-\sqrt{\Gamma_{bR/L}}$, which sets the phase shifts from \tabref{tab:ds}.
Then, the second condition implies that $\ket{S}$ and $\ket{T}$ must be uncoupled, i.e., that $\omegaA'=\omegaB'$ and $g=g^*$.
These constraints, in turn, imply that $\omegaA=\omegaB$ and $\gamma_R=\gamma_L$ for all setups, with the exception of nested giant atoms.
The symmetry of nested atoms with the dark-state phase shifts allows $g=g^*$ without the need to impose bidirectional coupling (see \tabref{tab:me_ctes} with $\phi_1=\phi_3=\{0,\,\pi\}$).
Note that the instances presented in \tabref{tab:ds} assume the bare relaxation rate to be the same at all connection points; however, conditions for arbitrary decay rates can be obtained using the same procedure.
 
With the Hamiltonian and the collapse operators in this basis [Eqs.~(\ref{eq:H_ST_app}) and (\ref{eq:L_ST_app}), respectively] and knowing that the master equation reads 
\begin{equation}
\dot{\rho} = -i\comm{H}{\rho} + \D[\COL_R]\rho + \D[\COL_L]\rho,
\end{equation}
we can derive the decay rates from the states $\ket{S}$ and $\ket{T}$ by tracking the time evolution of the ground-state population:
\begin{equation}
\begin{aligned}
    \dot{\rho}_{gggg} =& \mel{gg}{\dot{\rho}}{gg} =\\
    =&  \frac{1}{2}(\GammaA + \GammaB + \Gammacol + \Gammacol^*) \rho_{TT}\\
    & + \frac{1}{2}(\GammaA + \GammaB - \Gammacol - \Gammacol^*)\rho_{SS}\\
    & + \frac{1}{2}(\GammaB-\GammaA-\Gammacol+\Gammacol^*)\rho_{ST}\\
    & + \frac{1}{2}(\GammaB-\GammaA+\Gammacol-\Gammacol^*)\rho_{TS},
    \label{eq:rho_gg}
\end{aligned}
\end{equation}
where the first two terms yield the decay rates from the triplet and singlet states:
\begin{equation}
\begin{aligned}
    \Gamma_T &= \frac{1}{2}(\GammaA + \GammaB + \Gammacol + \Gammacol^*)\\
    \Gamma_S &= \frac{1}{2}(\GammaA + \GammaB - \Gammacol - \Gammacol^*).
\end{aligned}
\end{equation}

Using the conditions in \tabref{tab:ds}, we find that when $\gamma_1=\gamma_2=\gamma_3=\gamma_4$, the last two terms in \eqref{eq:rho_gg} cancel and that
\begin{equation}
    \begin{aligned}
        &\ket{D}=\ket{T} \implies \Gamma_T=0,\quad \Gamma_S=\GammaA+\GammaB\\
        &\ket{D} =\ket{S}\implies \Gamma_S=0,\quad \Gamma_T=\GammaA+\GammaB.
    \end{aligned}
\end{equation}
This means that whenever $\ket{D}=\ket{T}$, the triplet state does not decay, agreeing with the fact that it is a dark state, whereas the singlet state decays with a superradiant rate $\GammaA+\GammaB$, thus making it a bright state. Conversely, when the singlet state is a dark state, the triplet state is bright.

\subsection{Coherently driven system}
\label{app:dsd_driven}

When we drive the system coherently as described in \eqref{eq:drive2}, the collapse operators do not change, and the Hamiltonian in \eqref{eq:H_ST_app} gets an extra term:
\begin{equation}
\begin{aligned}   
    H_d =& -i\beta \SCA_R \COL_R^\dagger + \Hc =\\
        =& \Big[-i\beta \SCA_R \frac{\sqrt{2}}{2} \mleft(\sqrt{\Gamma_{aR}^*}-\sqrt{\Gamma_{bR}^*}\,\mright)
        \mleft(\ketbra{ee}{S}-\ketbra{S}{gg}\mright) \\
        & -i\beta \SCA_R \frac{\sqrt{2}}{2} \mleft(\sqrt{\Gamma_{aR}^*}+\sqrt{\Gamma_{bR}^*}\,\mright)
        \mleft(\ketbra{ee}{T}+\ketbra{T}{gg}\mright)\Big]\\
        &+\Hc
        \label{eq:Hd_eSTg}
\end{aligned}
\end{equation}
In this case, the possible dark states are
\begin{equation}
\ket{D_{S/T}} = \frac{1}{\sqrt{1 + \abs{\alpha}^2}} \mleft(\alpha\ket{S/T} + \ket{gg}\mright),
\end{equation}
as mentioned in \eqref{eq:d_st}.

Since the collapse operators have not changed, they impose the same dark-state restrictions as without the drive, i.e.,  $\ket{D}=\ket{D_S}$ when $\sqrt{\Gamma_{aR/L}}=\sqrt{\Gamma_{bR/L}}$ and $\ket{D}=\ket{D_T}$ when $\sqrt{\Gamma_{aR/L}}=-\sqrt{\Gamma_{bR/L}}$, which sets the phase shifts in \tabref{tab:dsd}.
These constraints, in turn, impose that the drive amplitudes of both atoms are equal ($\OmegaA=\OmegaB$) when $\ket{D_S}$ is dark and opposite ($\OmegaA=-\OmegaB$) when $\ket{D_T}$ is dark. In both cases, we define $\Omega\equiv\OmegaA$.

The second dark-state condition, which requires $\ket{D}$ to be an eigenstate of the Hamiltonian $H_u+H_d$, yields
\begin{equation}
    \alpha=\dfrac{i\sqrt{2}\Omega}{2\xi},
\end{equation}
with the expression of $\alpha$ as a function of the parameters of the system collected in \tabref{tab:dsd} for all the atomic setups. The second condition also implies that the existence of the dark state in this regime is subject to the atoms having opposite detunings, i.e., $\deltaA=-\deltaB$.

It is now convenient to change to the basis $\{\ket{D_{S/T}},\, \ket{B_{S/T}},\, \ket{T/S},\, \ket{ee}\}$, with $\ket{B_{S/T}}$ shown in \eqref{eq:b_st}. In this way, we can monitor the time evolution of the occupancy of the state $\ket{D_{S/T}}$ and derive at which rate $\GammaD$ this state is being populated. In this basis, the undriven Hamiltonian and the collapse operators read

\begin{widetext}
\begin{align}
    H_u =& +\frac{\omegaA'+\omegaB'}{2}\ketbra{ee} 
\pm \frac{g+g^*}{2}\ketbra{T/S} -\frac{(\omegaA'+\omegaB')\pm (g+g^*)\abs{\alpha}^2}{2\mleft(1+\abs{\alpha}^2\mright)}\ketbra{D_{S/T}}\nn\\
    &- \frac{(\omegaA'+\omegaB')\abs{\alpha}^2 \pm (g+g^*)}{2\mleft(1+\abs{\alpha}^2\mright)} \ketbra{B_{S/T}} + \Bigg[\alpha\frac{(\omegaA'-\omegaB') \pm (g-g^*)}{2\sqrt{1+\abs{\alpha}^2}}\ketbra{T/S}{D_{S/T}}\\ 
    &+\frac{(\omegaA'-\omegaB') \pm (g-g^*)}{2\sqrt{1+\abs{\alpha}^2}}\ketbra{T/S}{B_{S/T}} 
    +\alpha\frac{(\omegaA'+\omegaB')\mp (g+ g^*)}{2\mleft(1+\abs{\alpha}^2\mright)}\ketbra{B_{S/T}}{D_{S/T}}\Bigg] + \Hc,\nn\\[10pt]
    \COL_{R/L} =
    & +\frac{\sqrt{2}}{2}\mleft(\sqrt{\Gamma_{bR/L}} \pm \sqrt{\Gamma_{aR/L}}\,\mright) \ketbra{T/S}{ee}\nn\\
    & +\frac{\sqrt{2}}{2\sqrt{1+\abs{\alpha}^2}}\mleft(\sqrt{\Gamma_{bR/L}} \mp \sqrt{\Gamma_{aR/L}}\,\mright) \Big(\alpha^*\ketbra{D_{S/T}}{ee} + \ketbra{B_{S/T}}{ee}\Big)\nn\\
    &+ \frac{\sqrt{2}}{2\sqrt{1+\abs{\alpha}^2}}\mleft(\sqrt{\Gamma_{aR/L}} \pm \sqrt{\Gamma_{bR/L}}\,\mright) \Big(\ketbra{D_{S/T}}{T/S} -\alpha\ketbra{B_{S/T}}{T/S}\Big)\\
    & + \frac{\sqrt{2}}{2\mleft(1+\abs{\alpha}^2\mright)}\mleft(\sqrt{\Gamma_{aR/L}} \mp \sqrt{\Gamma_{bR/L}}\,\mright) \Big(\alpha\ketbra{D_{S/T}} -\alpha\ketbra{B_{S/T}}\nn\\ &\hspace{6.2cm}+\ketbra{D_{S/T}}{B_{S/T}}-\alpha^2\ketbra{B_{S/T}}{D_{S/T}}\Big), \nn
\end{align}
\end{widetext}
where $\pm$ ($\mp$) means that we take $+$ ($-$) in the basis $\{\ket{D_S},\, \ket{B_S},\, \ket{T},\, \ket{ee}\}$ and $-$ ($+$) in the basis $\{\ket{D_T},\, \ket{B_T},\, \ket{S},\, \ket{ee}\}$. The drive Hamiltonian is $H_d = -i\beta \SCA_R\COL_R^\dagger + \Hc$ with the above definition of $\COL_R$.

Similarly to the previous section, we now have all the elements necessary to derive the master equation. In particular, we find that the dark and bright state evolve as follows:
\begin{align}
    \label{eq:rho_d_st}
    \dot{\rho}_{D_{S/T},\,D_{S/T}} =& \frac{\GammaA+\GammaB\pm \Gammacol\pm \Gammacol^*}{2\mleft(1+\abs{\alpha}^2\mright)}\;\rho_{T/S,\,T/S} \nn\\
    & + \dots\\
    \dot{\rho}_{B_{S/T},\,B_{S/T}} =& \frac{\abs{\alpha}^2(\GammaA+\GammaB\pm \Gammacol\pm \Gammacol^*)}{2\mleft(1+\abs{\alpha}^2\mright)}\;\rho_{T/S,\,T/S} \nn\\
    & + \dots
    \label{eq:rho_b_st}
\end{align}
where we are only interested in tracking the dependence on the population of the $\ket{T/S}$ state. From \eqref{eq:rho_d_st}, we can define the effective pumping rate of the dark state:
\begin{equation}
    \GammaD = \frac{\GammaA+\GammaB\pm \Gammacol\pm \Gammacol^*}{2\mleft(1+\abs{\alpha}^2\mright)}.
\end{equation}

Using the conditions from \tabref{tab:dsd}, we find that when $\gamma_1=\gamma_2=\gamma_3=\gamma_4$, the dark state is populated at a rate
\begin{equation}
    \GammaD=\frac{\GammaA+\GammaB}{1+\abs{\alpha}^2},
\end{equation}
with the full expression of $\GammaD$ for each layout in \tabref{tab:gamma}. Similarly, the bright state is populated at a rate $\abs{\alpha}^2\GammaD$.

We note that the procedure used in this appendix can also be applied to multiple atoms with multiple connection points by using the pertinent Hamiltonian and collapse operators. 

\bibliography{References.bib}

\end{document}

%% file: Small.tex
\begin{tikzpicture}
\tikzset{every node}=[font=\sffamily]

\draw[white] (-3.5,-2.75) rectangle (4,2);

\draw[rounded corners] (-3.5,-2.75) rectangle (3.5,2.2);
\node[anchor = north west] at (-3.5, 2.15) {\textbf{(a) Small}};

\draw[ultra thick] (-3, 0) -- (3, 0);

\draw[very thick, MidnightBlue] (-1, 1) circle (15pt);
\node[MidnightBlue] at (-1, 1) {$a$};
\draw[very thick, MidnightBlue] (-1.3, 1.2) -- (-0.7, 1.2);
\draw[very thick, MidnightBlue] (-1.3, 0.8) -- (-0.7, 0.8);
\node[MidnightBlue] at (-1.8, 1.25) {$\ket{e}$};
\node[MidnightBlue] at (-1.8, 0.75) {$\ket{g}$};
\draw[MidnightBlue, <->] (-0.3, 0.8) -- (-0.3, 1.2);
\node[MidnightBlue] at (0, 1) {$\omegaA$};

\fill[MidnightBlue] (-1, 0) circle (3pt);
\draw[very thick, MidnightBlue] (-1, 0) -- (-1, 0.49);
\draw (-0.95, -0.15) -- (-0.95, -0.25);
\draw[->] (-0.95, -0.25) -- (-0.5, -0.25);
\node at (-0.65, -0.45) {$\gamma_{1R}$};
\draw (-1.05, -0.15) -- (-1.05, -0.25);
\draw[->] (-1.05, -0.25) -- (-1.5, -0.25);
\node at (-1.3, -0.45) {$\gamma_{1L}$};
\draw[thick, -latex] (-3.2, -0.25) -- (-2.5, -0.25);
\node at (-2.85, -0.5) {$\beta$};

\draw[very thick, MidnightBlue] (1, -1) circle (15pt);
\node[MidnightBlue] at (1, -1) {$b$};
\draw[very thick, MidnightBlue] (1.3, -1.2) -- (0.7, -1.2);
\draw[very thick, MidnightBlue] (1.3, -0.8) -- (0.7, -0.8);
\draw[MidnightBlue, <->] (1.7, -0.8) -- (1.7, -1.2);
\node[MidnightBlue] at (2, -1) {$\omegaB$};

\fill[MidnightBlue] (1, 0) circle (3pt);
\draw[very thick, MidnightBlue] (1, 0) -- (1, -0.49);
\draw (0.95, 0.15) -- (0.95, 0.25);
\draw[->] (0.95, 0.25) -- (0.5, 0.25);
\node at (0.65, 0.45) {$\gamma_{2L}$};
\draw (1.05, 0.15) -- (1.05, 0.25);
\draw[->] (1.05, 0.25) -- (1.5, 0.25);
\node at (1.3, 0.45) {$\gamma_{2R}$};

\draw (-3, -2) -- (3, -2);
\draw (-1, -2.1) -- (-1, -1.9);
\draw (1, -2.1) -- (1, -1.9);
\node at (0, -2.2) {$\phi$};

\end{tikzpicture}

%% file: Separate.tex
\begin{tikzpicture}
\tikzset{every node}=[font=\small\sffamily]

\draw[rounded corners] (-3,-2.75) rectangle (4,2.2);
\node[anchor = north west] at (-3, 2.15) {\textbf{(b) Separate}};


\draw[ultra thick] (-2.5, 0) -- (3.5, 0);

\draw[very thick, MidnightBlue] (-1, 1) circle (15pt);
\node[MidnightBlue] at (-1, 1) {$a$};
\draw[very thick, MidnightBlue] (-1.3, 1.2) -- (-0.7, 1.2);
\draw[very thick, MidnightBlue] (-1.3, 0.8) -- (-0.7, 0.8);
\fill[MidnightBlue] (-2, 0) circle (3pt);
\draw[very thick, MidnightBlue] (-2, 0) -- (-2, 1.02);
\draw[very thick, MidnightBlue] (-2.01, 1) -- (-1.53, 1);
\draw (-1.95, -0.15) -- (-1.95, -0.25);
\draw[->] (-1.95, -0.25) -- (-1.5, -0.25);
\node at (-1.65, -0.45) {$\gamma_{1R}$};
\draw (-2.05, -0.15) -- (-2.05, -0.25);
\draw[->] (-2.05, -0.25) -- (-2.5, -0.25);
\node at (-2.3, -0.45) {$\gamma_{1L}$};
\fill[MidnightBlue] (0, 0) circle (3pt);
\draw[very thick, MidnightBlue] (0, 0) -- (0, 1.02);
\draw[very thick, MidnightBlue] (0.01, 1) -- (-0.48, 1);
\draw (0.05, -0.15) -- (0.05, -0.25);
\draw[->] (0.05, -0.25) -- (0.5, -0.25);
\node at (0.35, -0.45) {$\gamma_{2R}$};
\draw (-0.05, -0.15) -- (-0.05, -0.25);
\draw[->] (-0.05, -0.25) -- (-0.5, -0.25);
\node at (-0.3, -0.45) {$\gamma_{2L}$};

\draw[very thick, MidnightBlue] (2,-1) circle (15pt);
\node[MidnightBlue] at (2,-1) {$b$};
\draw[very thick, MidnightBlue] (2.3, -1.2) -- (1.7, -1.2);
\draw[very thick, MidnightBlue] (2.3, -0.8) -- (1.7, -0.8);
\fill[MidnightBlue] (1, 0) circle (3pt);
\draw[very thick, MidnightBlue] (1, 0) -- (1, -1.02);
\draw[very thick, MidnightBlue] (0.99, -1) -- (1.47, -1);
\draw (0.95, 0.15) -- (0.95, 0.25);
\draw[->] (0.95, 0.25) -- (0.5, 0.25);
\node at (0.65, 0.45) {$\gamma_{3L}$};
\draw (1.05, 0.15) -- (1.05, 0.25);
\draw[->] (1.05, 0.25) -- (1.5, 0.25);
\node at (1.3, 0.45) {$\gamma_{3R}$};
\fill[MidnightBlue] (3, 0) circle (3pt);
\draw[very thick, MidnightBlue] (3, 0) -- (3, -1.02);
\draw[very thick, MidnightBlue] (3.01, -1) -- (2.52, -1.0);
\draw (2.95, 0.15) -- (2.95, 0.25);
\draw[->] (2.95, 0.25) -- (2.5, 0.25);
\node at (2.65, 0.45) {$\gamma_{4L}$};
\draw (3.05, 0.15) -- (3.05, 0.25);
\draw[->] (3.05, 0.25) -- (3.5, 0.25);
\node at (3.3, 0.45) {$\gamma_{4R}$};

\draw (-2.5, -2) -- (3.5, -2);
\draw (-2, -2.1) -- (-2, -1.9);
\draw (0, -2.1) -- (0, -1.9);
\draw (1, -2.1) -- (1, -1.9);
\draw (3, -2.1) -- (3, -1.9);
\node at (-1, -2.2) {$\phi_1$};
\node at (0.5, -2.2) {$\phi_2$};
\node at (2, -2.2) {$\phi_3$};

\end{tikzpicture}

%% file: Nested.tex
\begin{tikzpicture}
\tikzset{every node}=[font=\small\sffamily]

\draw[white] (-3.5,-2.75) rectangle (4,2.5);

\draw[rounded corners] (-3.5,-2.75) rectangle (3.5,2.2);
\node[anchor = north west] at (-3.5, 2.15) {\textbf{(c) Nested}};

\draw[ultra thick] (-3, 0) -- (3, 0);

\draw[very thick, MidnightBlue] (0, 1) circle (15pt);
\node[MidnightBlue] at (0, 1) {$a$};
\draw[very thick, MidnightBlue] (-0.3, 1.2) -- (0.3, 1.2);
\draw[very thick, MidnightBlue] (-0.3, 0.8) -- (0.3, 0.8);
\fill[MidnightBlue] (-2.5, 0) circle (3pt);
\draw[very thick, MidnightBlue] (-2.5, 0) -- (-2.5, 1.02);
\draw[very thick, MidnightBlue] (-2.5, 1) -- (-0.53, 1);
\draw (-2.45, -0.15) -- (-2.45, -0.25);
\draw[->] (-2.45, -0.25) -- (-2, -0.25);
\node at (-2.15, -0.45) {$\gamma_{1R}$};
\draw (-2.55, -0.15) -- (-2.55, -0.25);
\draw[->] (-2.55, -0.25) -- (-3, -0.25);
\node at (-2.8, -0.45) {$\gamma_{1L}$};
\fill[MidnightBlue] (2.5, 0) circle (3pt);
\draw[very thick, MidnightBlue] (2.5, 0) -- (2.5, 1.02);
\draw[very thick, MidnightBlue] (2.51, 1) -- (0.52, 1.0);
\draw (2.45, -0.15) -- (2.45, -0.25);
\draw[->] (2.45, -0.25) -- (2, -0.25);
\node at (2.2, -0.45) {$\gamma_{4L}$};
\draw (2.55, -0.15) -- (2.55, -0.25);
\draw[->] (2.55, -0.25) -- (3, -0.25);
\node at (2.85, -0.45) {$\gamma_{4R}$};

\draw[very thick, MidnightBlue] (0, -1) circle (15pt);
\node[MidnightBlue] at (0, -1) {$b$};
\draw[very thick, MidnightBlue] (-0.3, -1.2) -- (0.3, -1.2);
\draw[very thick, MidnightBlue] (-0.3, -0.8) -- (0.3, -0.8);
\fill[MidnightBlue] (-1, 0) circle (3pt);
\draw[very thick, MidnightBlue] (-1, 0) -- (-1, -1.02);
\draw[very thick, MidnightBlue] (-1.01, -1) -- (-0.52, -1);
\draw (-0.95, 0.15) -- (-0.95, 0.25);
\draw[->] (-0.95, 0.25) -- (-0.5, 0.25);
\node at (-0.7, 0.45) {$\gamma_{2R}$};
\draw (-1.05, 0.15) -- (-1.05, 0.25);
\draw[->] (-1.05, 0.25) -- (-1.5, 0.25);
\node at (-1.3, 0.45) {$\gamma_{2L}$};
\fill[MidnightBlue] (1, 0) circle (3pt);
\draw[very thick, MidnightBlue] (1, 0) -- (1, -1.02);
\draw[very thick, MidnightBlue] (0.99, -1) -- (0.52, -1);
\draw (0.95, 0.15) -- (0.95, 0.25);
\draw[->] (0.95, 0.25) -- (0.5, 0.25);
\node at (0.7, 0.45) {$\gamma_{3L}$};
\draw (1.05, 0.15) -- (1.05, 0.25);
\draw[->] (1.05, 0.25) -- (1.5, 0.25);
\node at (1.3, 0.45) {$\gamma_{3R}$};

\draw (-3, -2) -- (3, -2);
\draw (-2.5, -2.1) -- (-2.5, -1.9);
\draw (-1, -2.1) -- (-1, -1.9);
\draw (1, -2.1) -- (1, -1.9);
\draw (2.5, -2.1) -- (2.5, -1.9);
\node at (-1.75, -2.2) {$\phi_1$};
\node at (0, -2.2) {$\phi_2$};
\node at (1.75, -2.2) {$\phi_3$};

\end{tikzpicture}

%% file: Braided.tex
\begin{tikzpicture}
\tikzset{every node}=[font=\small\sffamily]

\draw[rounded corners] (-3.5,-2.75) rectangle (3.5,2.2);
\node[anchor = north west] at (-3.5, 2.15) {\textbf{(d) Braided}};

\draw[ultra thick] (-3, 0) -- (3, 0);

\draw[very thick, MidnightBlue] (0, 1) circle (15pt);
\node[MidnightBlue] at (0, 1) {$a$};
\draw[very thick, MidnightBlue] (-0.3, 1.2) -- (0.3, 1.2);
\draw[very thick, MidnightBlue] (-0.3, 0.8) -- (0.3, 0.8);
\fill[MidnightBlue] (-2.5, 0) circle (3pt);
\draw[very thick, MidnightBlue] (-2.5, 0) -- (-2.5, 1.02);
\draw[very thick, MidnightBlue] (-2.5, 1) -- (-0.53, 1);
\draw (-2.45, -0.15) -- (-2.45, -0.25);
\draw[->] (-2.45, -0.25) -- (-2, -0.25);
\node at (-2.15, -0.45) {$\gamma_{1R}$};
\draw (-2.55, -0.15) -- (-2.55, -0.25);
\draw[->] (-2.55, -0.25) -- (-3, -0.25);
\node at (-2.8, -0.45) {$\gamma_{1L}$};
\fill[MidnightBlue] (1, 0) circle (3pt);
\draw[very thick, MidnightBlue] (1, 0) -- (1, 1.02);
\draw[very thick, MidnightBlue] (0.99, 1) -- (0.52, 1);
\draw (0.95, -0.15) -- (0.95, -0.25);
\draw[->] (0.95, -0.25) -- (0.5, -0.25);
\node at (0.7, -0.45) {$\gamma_{3L}$};
\draw (1.05, -0.15) -- (1.05, -0.25);
\draw[->] (1.05, -0.25) -- (1.5, -0.25);
\node at (1.3, -0.45) {$\gamma_{3R}$};

\draw[very thick, MidnightBlue] (0, -1) circle (15pt);
\node[MidnightBlue] at (0, -1) {$b$};
\draw[very thick, MidnightBlue] (-0.3, -1.2) -- (0.3, -1.2);
\draw[very thick, MidnightBlue] (-0.3, -0.8) -- (0.3, -0.8);
\fill[MidnightBlue] (-1, 0) circle (3pt);
\draw[very thick, MidnightBlue] (-1, 0) -- (-1, -1.02);
\draw[very thick, MidnightBlue] (-1.01, -1) -- (-0.52, -1);
\draw (-0.95, 0.15) -- (-0.95, 0.25);
\draw[->] (-0.95, 0.25) -- (-0.5, 0.25);
\node at (-0.7, 0.45) {$\gamma_{2R}$};
\draw (-1.05, 0.15) -- (-1.05, 0.25);
\draw[->] (-1.05, 0.25) -- (-1.5, 0.25);
\node at (-1.3, 0.45) {$\gamma_{2L}$};

\fill[MidnightBlue] (2.5, 0) circle (3pt);
\draw[very thick, MidnightBlue] (2.5, 0) -- (2.5, -1.02);
\draw[very thick, MidnightBlue] (2.51, -1) -- (0.52, -1.0);
\draw (2.45, 0.15) -- (2.45, 0.25);
\draw[->] (2.45, 0.25) -- (2, 0.25);
\node at (2.2, 0.45) {$\gamma_{4L}$};
\draw (2.55, 0.15) -- (2.55, 0.25);
\draw[->] (2.55, 0.25) -- (3, 0.25);
\node at (2.85, 0.45) {$\gamma_{4R}$};

\draw (-3, -2) -- (3, -2);
\draw (-2.5, -2.1) -- (-2.5, -1.9);
\draw (-1, -2.1) -- (-1, -1.9);
\draw (1, -2.1) -- (1, -1.9);
\draw (2.5, -2.1) -- (2.5, -1.9);
\node at (-1.75, -2.2) {$\phi_1$};
\node at (0, -2.2) {$\phi_2$};
\node at (1.75, -2.2) {$\phi_3$};

\end{tikzpicture}

%% file: TableME.tex
\begin{table*}[t]
\caption{Frequency shifts, exchange interaction, individual and collective decay rates [$\delta\omega_j, g, \Gamma_j, \Gammacol$ in \eqref{eq:me}] for two small and giant atoms chirally coupled to a 1D open waveguide. We assume arbitrary phase shifts $\phi_1,\phi_2,\phi_3$ and equal bare relaxation rates at each coupling point $k$, i.e., $ \gamma_R\equiv\gamma_{kR}$ and $\gamma_L\equiv\gamma_{kL}$ for $k=1,2,3,4$. For the general case with different $\gamma_k$, see \tabref{tab:me_ctes_full} in \appref{app:me}.}
\label{tab:me_ctes}
\centering
\setlength{\extrarowheight}{8pt}
\setlength{\tabcolsep}{10pt}
\begin{tabular}{c c l l}
\hline\hline
Coefficient & Topology & Atom $a$ & Atom $b$\\[4pt]\hline
\multirow{4}{*}{\begin{tabular}{c}\textbf{Frequency shifts,}\\ $\delta\omegaA,\, \delta\omegaB$\end{tabular}} & Small  & 0 & 0\\
&Separate  & $(\gamma_{R}+\gamma_{L})\sin\phi_1$ & 	$(\gamma_{R}+\gamma_{L})\sin\phi_3$\\
&Nested  & $(\gamma_{R}+\gamma_{L})\sin(\phi_1+\phi_2+	\phi_3)$ & $(\gamma_{R}+\gamma_{L})\sin\phi_2$\\
&Braided & $(\gamma_{R}+\gamma_{L})\sin(\phi_1+\phi_2)$ & $(\gamma_{R}+\gamma_{L})\sin(\phi_2+\phi_3)$\\[4pt]
\hline
\multirow{4}{*}{\begin{tabular}{c}\textbf{Individual decays,}\\ $\GammaA,\, \GammaB$\end{tabular}} & Small  & $\gamma_{R} + \gamma_{L}$ & $\gamma_{R} + \gamma_{L}$ \\
&Separate  &  $2(\gamma_{R} +\gamma_{L})(1+\cos\phi_1)$ & 	$2(\gamma_{R} +\gamma_{L})(1+\cos\phi_3)$\\
&Nested  & $2(\gamma_{R} +\gamma_{L})[1+\cos(\phi_1+\phi_2+\phi_3)]$ & 	$2(\gamma_{R} +\gamma_{L})(1+\cos\phi_2)$\\
&Braided & $2(\gamma_{R} +\gamma_{L})[1+\cos(\phi_1+\phi_2)]$ & 	$2(\gamma_{R} +\gamma_{L})[1+\cos(\phi_2+\phi_3)]$\\[4pt]
\hline
\multirow{6}{*}{\begin{tabular}{c}\textbf{Collective decay,}\\ $\Gammacol$\end{tabular}} & Small  & \multicolumn{2}{l}{$\gamma_{R}\,e^{i\phi}+\gamma_{L}\,e^{-i\phi}$}\\
&Separate  & \multicolumn{2}{l}{$\gamma_{R} (e^{i(\phi_1+\phi_2)} + e^{i(\phi_1+\phi_2+\phi_3)} + e^{i\phi_2} + e^{i(\phi_2+\phi_3)}) +$}\\[-6pt]
& 	& \multicolumn{2}{l}{$\gamma_{L}( e^{-i(\phi_1+\phi_2)} + e^{-i(\phi_1+	\phi_2+\phi_3)} + e^{-i\phi_2} + e^{-i(\phi_2+\phi_3)})$} \\
&Nested  & \multicolumn{2}{l}{$\gamma_{R} (e^{i\phi_1} +  e^{i(\phi_1+	\phi_2)} + e^{-i(\phi_2+\phi_3)} + e^{-i\phi_3}) +$}\\[-6pt]
&	& \multicolumn{2}{l}{$\gamma_{L} 	(e^{-i\phi_1} + e^{-i(\phi_1+	\phi_2)} + 		e^{i(\phi_2+\phi_3)} + e^{i\phi_3})$} \\
&Braided & \multicolumn{2}{l}{$\gamma_{R} (e^{i\phi_1} + e^{i(\phi_1+		\phi_2+\phi_3)} +  e^{-i\phi_2} + e^{i\phi_3}) +$}\\[-6pt]
& 	&  \multicolumn{2}{l}{$\gamma_{L} 	(e^{-i\phi_1} + e^{-i(\phi_1+\phi_2+\phi_3)} + 	e^{i\phi_2} + e^{-i\phi_3})$} \\[4pt]
\hline
\multirow{6}{*}{\begin{tabular}{c}\textbf{Exchange}\\[-4pt] \textbf{interaction,}\\ $g$\end{tabular}} & Small  & \multicolumn{2}{l}{$[\gamma_{R}\,e^{i\phi}-\gamma_{L}\,e^{-i\phi}]/(2i)$}\\
& Separate  & \multicolumn{2}{l}{$[\gamma_{R} (e^{i(\phi_1+\phi_2)} + e^{i(\phi_1+\phi_2+\phi_3)} + e^{i\phi_2} + e^{i(\phi_2+\phi_3)}) -$}\\[-6pt]
&	& \multicolumn{2}{l}{$\gamma_{L}( e^{-i(\phi_1+\phi_2)} + e^{-i(\phi_1+\phi_2+\phi_3)} + e^{-i\phi_2} + e^{-i(\phi_2+\phi_3)})]/(2i)$} \\
&Nested  & \multicolumn{2}{l}{$[\gamma_{R} (e^{i\phi_1} +  e^{i(\phi_1+\phi_2)} - e^{-i(\phi_2+\phi_3)} - e^{-i\phi_3}) -$}\\[-6pt]
&	& \multicolumn{2}{l}{$\gamma_{L} (e^{-i\phi_1} + e^{-i(\phi_1+	\phi_2)} - 	e^{i(\phi_2+\phi_3)} - e^{i\phi_3})]/(2i)$} \\
&Braided & \multicolumn{2}{l}{$[\gamma_{R} (e^{i\phi_1} + e^{i(\phi_1+\phi_2+\phi_3)} -  e^{-i\phi_2} + e^{i\phi_3}) -$}\\[-6pt]
&	& \multicolumn{2}{l}{$\gamma_{L} (e^{-i\phi_1} + e^{-i(\phi_1+\phi_2+\phi_3)} - e^{i\phi_2} +e^{-i\phi_3})]/(2i)$} \\[4pt]
\hline\hline
\end{tabular}
\end{table*}

%% file: Basis.tex
\begin{tikzpicture}[decoration = {snake, pre length=1pt, post length=2pt,},]
\tikzset{every node}=[font=\small\sffamily]

\draw[rounded corners] (0,0) rectangle (5,4);
\node at (0.45, 3.65) {\textbf{(a)}};

\draw[ultra thick, black] (2, 3.5) -- (3, 3.5);
\node at (3.4 ,3.5) {$\ket{ee}$};

\draw[ultra thick, black] (1, 2) -- (2, 2);
\node at (0.6, 2) {$\ket{eg}$};

\draw[ultra thick, black] (3, 2) -- (4, 2);
\node at (4.4 , 2) {$\ket{ge}$};

\draw[ultra thick, black] (2, 0.5) -- (3, 0.5);
\node at (3.4, 0.5) {$\ket{gg}$};

\path[draw=MidnightBlue, thick, ->, decorate] (2.15, 3.3) -- (1.35, 2.2); 
\node[MidnightBlue] at (1.3, 2.9) {$\GammaB$};
\node[RedOrange] at (2.2, 2.4) {$\Omega_b$};
\path[draw=RedOrange, thick, <->] (2.45, 3.3) -- (1.65, 2.2); 

\path[draw=MidnightBlue, thick, ->, decorate] (2.85, 3.3) -- (3.65, 2.2); 
\node[MidnightBlue] at (3.7, 2.9) {$\GammaA$};
\node[RedOrange] at (2.8, 2.4) {$\Omega_a$};
\path[draw=RedOrange, thick, <->] (2.55, 3.3) -- (3.35, 2.2); 

\path[draw=MidnightBlue, thick, ->, decorate] (1.35, 1.8) -- (2.15, 0.7); 
\node[MidnightBlue] at (1.3, 1.2) {$\GammaA$};
\node[RedOrange] at (2.2, 1.6) {$\Omega_a$};
\path[draw=RedOrange, thick, <->] (1.65, 1.8) -- (2.45, 0.7); 

\path[draw=MidnightBlue, thick, ->, decorate] (3.65, 1.8) -- (2.85, 0.7); 
\node[MidnightBlue] at (3.7, 1.2) {$\GammaB$};
\node[RedOrange] at (2.8, 1.6) {$\Omega_b$};
\path[draw=RedOrange, thick, <->] (3.35, 1.8) -- (2.55, 0.7); 

\draw[rounded corners] (5.5,0) rectangle (10.5,4);
\node at (5.85, 3.65) {\textbf{(b)}};

\draw[ultra thick, black] (8.5, 3.5) -- (9.5, 3.5);
\node at (9.9 ,3.5) {$\ket{ee}$};

\draw[ultra thick, black] (6.5, 2) -- (7.5, 2);
\node at (6, 2) {$\ket{S/T}$};

\draw[ultra thick, black] (8.5, 2) -- (9.5, 2);
\node at (10 , 2) {$\ket{T/S}$};

\draw[ultra thick, black] (8.5, 0.5) -- (9.5, 0.5);
\node at (9.9, 0.5) {$\ket{gg}$};

\draw[<->] (7.7, 2) -- (8.3, 2);
\node at (8, 2.3) {$\xi$};

\path[draw=MidnightBlue, thick, ->, decorate] (8.85, 3.3) -- (8.85, 2.2); 
\node[MidnightBlue] at (8, 2.9) {$\GammaA+\GammaB$};
\path[draw=RedOrange, thick, <->] (9.15, 3.3) -- (9.15, 2.2); 
\node[RedOrange] at (9.7, 2.9) {$\sqrt{2}\Omega$};

\path[draw=MidnightBlue, thick, ->, decorate] (8.85, 1.8) -- (8.85, 0.7); 
\node[MidnightBlue] at (8, 1.2) {$\GammaA+\GammaB$};
\path[draw=RedOrange, thick, <->] (9.15, 1.8) -- (9.15, 0.7); 
\node[RedOrange] at (9.7, 1.2) {$\sqrt{2}\Omega$};

\draw[rounded corners] (11,0) rectangle (16,4);
\node at (11.35, 3.65) {\textbf{(c)}};

\draw[ultra thick, black] (13, 3.5) -- (14, 3.5);
\node at (14.4 ,3.5) {$\ket{ee}$};

\draw[ultra thick, black] (13, 2) -- (14, 2);
\node at (14.5, 2) {$\ket{T/S}$};

\draw[ultra thick, black] (12.2, 0.5) -- (13.2, 0.5);
\node at (11.6, 0.5) {$\ket{D_{S/T}}$};

\draw[ultra thick, black] (13.8, 0.5) -- (14.8, 0.5);
\node at (15.4 , 0.5) {$\ket{B_{S/T}}$};

\path[draw=MidnightBlue, thick, ->, decorate] (13.35, 3.3) -- (13.35, 2.2); 
\node[MidnightBlue] at (12.5, 2.9) {$\GammaA+\GammaB$};
\path[draw=RedOrange, thick, <->] (13.65, 3.3) -- (13.65, 2.2); 
\node[RedOrange] at (14.2, 2.9) {$\sqrt{2}\Omega$};

\path[draw=MidnightBlue, thick, ->, decorate] (13.3, 1.8) -- (12.7, 0.7); 
\node[MidnightBlue] at (12.4, 1.3) {$\GammaD$};

\path[draw=MidnightBlue, thick, ->, decorate] (13.8, 1.8) -- (14.4, 0.7); 
\node[MidnightBlue] at (15, 1.3) {$\abs{\alpha}^2\GammaD$};
\path[draw=RedOrange, thick, <->] (13.6, 1.8) -- (14.1, 0.7); 

\end{tikzpicture}

%% file: TableDS.tex
\begin{table}[t]
\caption{Conditions for the existence of a nontrivial dark state $\ket{D}$ in the setups described in \figref{fig:setups}. We assume arbitrary phase shifts $\phi_1,\phi_2,\phi_3$ and equal bare relaxation rates at each coupling point $k$, i.e., $ \gamma_R\equiv\gamma_{kR}$ and $\gamma_L\equiv\gamma_{kL}$ for $k=1,2,3,4$. In all cases, the frequencies of atom $a$ and $b$ must be equal ($\omegaA=\omegaB$).}
\label{tab:ds}
\centering
\setlength{\extrarowheight}{6pt}
\begin{tabularx}{\linewidth}{C C C C}
\hline\hline
Topology & Dark state & Phase shift  & Chirality \tabularnewline[4pt]\hline
\multirow{2.3}{*}{Small} & $ \ket{S}$ & $\phi=0$ & $\gamma_R=\gamma_L$ \tabularnewline
&$ \ket{T}$ & $\phi=\pi$ & $\gamma_R=\gamma_L$ \tabularnewline[4pt]\hline
\multirow{3.5}{*}{Separate} & \multirow{2}{*}{$ \ket{S}$} & $\phi_1=\phi_3\neq\pi$   & \multirow{2}{*}{$\gamma_R=\gamma_L$} \tabularnewline[-6pt]
&&$\phi_1=-\phi_2$&\tabularnewline
& \multirow{2}{*}{$ \ket{T}$} & $\phi_1=\phi_3\neq\pi$   & \multirow{2}{*}{$\gamma_R=\gamma_L$}\tabularnewline[-6pt]
&&$\phi_1+\phi_2 = \pi$&\tabularnewline[4pt]\hline
\multirow{3.5}{*}{Nested} & \multirow{2}{*}{$ \ket{S}$} & $\phi_1=\phi_3=0$   & \multirow{2}{*}{Any} \tabularnewline[-6pt]
&&$\phi_2\neq\pi$&\tabularnewline
& \multirow{2}{*}{$ \ket{T}$} & $\phi_1=\phi_3=\pi$   & \multirow{2}{*}{Any}\tabularnewline[-6pt]
&&$\phi_2 \neq \pi$&\tabularnewline[4pt]\hline
\multirow{3.5}{*}{Braided} & \multirow{2}{*}{$ \ket{S}$} & $\phi_1=\phi_3=0$   & \multirow{2}{*}{$\gamma_R=\gamma_L$} \tabularnewline[-6pt]
&&$\phi_2\neq\pi$&\tabularnewline
& \multirow{2}{*}{$ \ket{T}$} & $\phi_1=\phi_3=\pi$   & \multirow{2}{*}{$\gamma_R=\gamma_L$}\tabularnewline[-6pt]
&&$\phi_2 \neq 0$&\tabularnewline[4pt]
\hline\hline
\end{tabularx}
\end{table}

%% file: TableDSd.tex
\begin{table}[t]
\caption{Conditions for the existence of a nontrivial dark state in the coherently driven versions of the setups in \figref{fig:setups}. $\ket{D_S}$ and $\ket{D_T}$ are the states defined in \eqref{eq:d_st}. We assume arbitrary phase shifts $\phi_1,\phi_2,\phi_3$ and equal bare relaxation rates at each coupling point $k$, i.e., $ \gamma_R\equiv\gamma_{kR}$ and $\gamma_L\equiv\gamma_{kL}$ for $k=1,2,3,4$. These dark states exist regardless of the chirality of the coupling between the atoms and the waveguide, but only when the atoms have opposite detunings ($\deltaA=-\deltaB$). We use the definitions $\delta\equiv\deltaA$, $\Omega\equiv\OmegaA$ and  $\Delta\gamma\equiv\gamma_L-\gamma_R$. Note that with the phase shifts below, $\OmegaA=\OmegaB$ when $\ket{D_S}$ is dark and $\OmegaA=-\OmegaB$ when $\ket{D_T}$ is dark.}
\label{tab:dsd}
\centering
\setlength{\extrarowheight}{6pt}
\begin{tabularx}{\linewidth}{>{\setlength\hsize{0.85\hsize}}C >{\setlength\hsize{0.6\hsize}}C >{\setlength\hsize{0.75\hsize}}C  >{\setlength\hsize{1.75\hsize}}C >{\setlength\hsize{1.05\hsize}}C}
\hline\hline
\multirow{1.75}{*}{Topology} &  Dark &  Phase & \multirow{1.75}{*}{$\alpha(\beta)$} & \multirow{1.75}{*}{$\alpha(\Omega)$} \tabularnewline[-6pt]
 &  state &  shift && \tabularnewline[4pt]
\hline
\multirow{3}{*}{Small} & $\ket{D_S}$ & $\phi=0$ & $\dfrac{2\sqrt{2}\beta\sqrt{\gamma_R}}{\Delta\gamma-2i\delta}$ & $ \dfrac{\sqrt{2}\Omega}{\Delta\gamma-2i\delta}$  \tabularnewline[12pt]
&$\ket{D_T}$ & $\phi=\pi$ &  $\dfrac{2\sqrt{2}\beta\sqrt{\gamma_R}}{\Delta\gamma-2i\delta}$ & $ \dfrac{\sqrt{2}\Omega}{\Delta\gamma-2i\delta}$  \tabularnewline[8pt]
\hline
\multirow{5}{*}{Separate} & \multirow{2.5}{*}{$\ket{D_S}$} & $\phi_1=0$ & \multirow{2.5}{*}{$\dfrac{2\sqrt{2}\beta\sqrt{\gamma_R}}{2\Delta\gamma-i\delta}$} & \multirow{2.5}{*}{$\dfrac{\sqrt{2}\Omega}{4\Delta\gamma-2i\delta}$} \tabularnewline[-6pt]
&&$\phi_2 = \pi$&&\tabularnewline[-6pt]
&&$\phi_3 = 0$&&\tabularnewline
& \multirow{2.5}{*}{$\ket{D_T}$} & $\phi_1=0$ & \multirow{2.5}{*}{$\dfrac{2\sqrt{2}\beta\sqrt{\gamma_R}}{2\Delta\gamma-i\delta}$} & \multirow{2.5}{*}{$ \dfrac{\sqrt{2}\Omega}{4\Delta\gamma-2i\delta}$} \tabularnewline[-6pt]
&&$\phi_2 = \pi$&&\tabularnewline[-6pt]
&&$\phi_3 = 0$&& \tabularnewline[4pt]
\hline
\multirow{5}{*}{Nested} & \multirow{2.5}{*}{$\ket{D_S}$} & $\phi_1=0$  & \multirow{2.5}{*}{$\dfrac{\sqrt{2}\beta\sqrt{\gamma_R}(1+e^{i\phi_2})}{-i\delta}$} & \multirow{2.5}{*}{$\dfrac{\sqrt{2}\Omega}{-2i\delta}$}   \tabularnewline[-6pt]
&&$\phi_2\neq\pi$ &&\tabularnewline[-6pt]
&&$\phi_3=0$ &&\tabularnewline
& \multirow{2.5}{*}{$\ket{D_T}$}& $\phi_1=\pi$ &  \multirow{2.5}{*}{$\dfrac{\sqrt{2}\beta\sqrt{\gamma_R}(1+e^{i\phi_2})}{-i\delta}$} & \multirow{2.5}{*}{$ \dfrac{\sqrt{2}\Omega}{-2i\delta}$}   \tabularnewline[-6pt]
&&$\phi_2 \neq \pi$ && \tabularnewline[-6pt]
&&$\phi_3 = \pi$ && \tabularnewline[4pt]
\hline
\multirow{5}{*}{Braided} & \multirow{2.5}{*}{$\ket{D_S}$} & $\phi_1=0$ & \multirow{2.5}{*}{$\dfrac{\sqrt{2}\beta\sqrt{\gamma_R}(1+e^{i\phi_2})}{\Delta\gamma-i\delta}$} & \multirow{2.5}{*}{$\dfrac{\sqrt{2}\Omega}{2\Delta\gamma-2i\delta}$} \tabularnewline[-6pt]
&&$\phi_2\neq\pi$ && \tabularnewline[-6pt]
&&$\phi_3=0$ && \tabularnewline
& \multirow{2.5}{*}{$\ket{D_T}$} & $\phi_1=\pi$ & \multirow{2.5}{*}{$\dfrac{\sqrt{2}\beta\sqrt{\gamma_R}(1-e^{i\phi_2})}{\Delta\gamma-i\delta}$} & \multirow{2.5}{*}{$\dfrac{\sqrt{2}\Omega}{2\Delta\gamma-2i\delta}$} \tabularnewline[-6pt]
&&$\phi_2 \neq 0$ && \tabularnewline[-6pt]
&&$\phi_3 = \pi$ &&\tabularnewline[4pt]
\hline\hline
\end{tabularx} 
\end{table}

%% file: TableGamma.tex
\begin{table*}[t]
\caption{Populating rate $\GammaD$ of the dark state in the driven-dissipative regime for all the setups described in \figref{fig:setups}. To calculate $\GammaD=(\GammaA+\GammaB)/(1+\abs{\alpha}^2)$, we take $\GammaA$ and $\GammaB$ from \tabref{tab:me_ctes} with the phase shifts from \tabref{tab:dsd}. We also take $\alpha$ from \tabref{tab:dsd} and assume equal bare relaxation rates at each coupling point $k$, i.e., $ \gamma_R\equiv\gamma_{kR}$ and $\gamma_L\equiv\gamma_{kL}$ for $k=1,2,3,4$. We use the definitions $\delta\equiv\deltaA=-\deltaB$, $\Omega\equiv\OmegaA=\pm\OmegaB$ and  $\Delta\gamma\equiv\gamma_L-\gamma_R$. In the expression of $\GammaD$ for braided giant atoms, $\pm$ denotes $+$ when the dark state is $\ket{D_S}$ and $-$ when it is $\ket{D_T}$.}
\label{tab:gamma}
\centering
\setlength{\extrarowheight}{8pt}
\setlength{\tabcolsep}{10pt}
\begin{tabular}{c c c}
\hline\hline
Topology & $\GammaD(\beta)$ & $\GammaD(\Omega)$ \\[4pt]
\hline
Small & $\dfrac{2(\gamma_R+\gamma_L)(\Delta\gamma^2 + 4\delta^2)}{8\gamma_R\abs{\beta}^2 + \Delta\gamma^2 + 4\delta^2}$ & 
$\dfrac{2(\gamma_R+\gamma_L)(\Delta\gamma^2 + 4\delta^2)}{2\Omega^2+\Delta\gamma^2 + 4\delta^2}$ \\[12pt]
Separate & $\dfrac{8(\gamma_R+\gamma_L)(4\Delta\gamma^2+\delta^2)}{8\gamma_R\abs{\beta}^2 + 4\Delta\gamma^2 + \delta^2}$ &
$\dfrac{16(\gamma_R+\gamma_L)(4\Delta\gamma^2+\delta^2)}{\Omega^2 + 8\Delta\gamma^2 + 2\delta^2}$ \\[12pt]
Nested & $\dfrac{4(\gamma_R+\gamma_L)(1+\cos\phi_2)\delta^2}{4\gamma_R\abs{\beta}^2(1+\cos\phi_2) + \delta^2}$ &
$\dfrac{8(\gamma_R+\gamma_L)(1+\cos\phi_2)\delta^2}{\Omega^2+2\delta^2}
$ \\[12pt]
Braided & $\dfrac{4(\gamma_R+\gamma_L)(1\pm\cos\phi_2)(\Delta\gamma^2+\delta^2)}{4\gamma_R\abs{\beta}^2(1\pm\cos\phi_2) + \Delta\gamma^2 + \delta^2}$ &
$\dfrac{8(\gamma_R+\gamma_L)(1\pm\cos\phi_2)(\Delta\gamma^2+\delta^2)}{\Omega^2+2\Delta\gamma^2+2\delta^2}
$ \\[12pt]
\hline\hline
\end{tabular} 
\end{table*}

%% file: Chain.tex
\begin{tikzpicture}
\tikzset{every node}=[font=\footnotesize\sffamily]

\draw[rounded corners] (-2,-1.25) rectangle (2.25,1.75);
\node[anchor = north west] at (-2, 1.75) {\textbf{(a) Braided chain}};

\draw[very thick] (-1.75, 0) -- (1.75, 0);
\draw[very thick, densely dotted] (1.8, 0) -- (2, 0);

\fill[MidnightBlue] (-1.5, 0) circle (2pt);
\fill[MidnightBlue] (-0.5, 0) circle (2pt);
\draw[ thick, MidnightBlue] (-1.5, 0) |- (-1.25, 0.7);
\draw[ thick, MidnightBlue] (-0.5, 0) |- (-0.75, 0.7);
\draw[ thick, MidnightBlue] (-1, 0.7) circle (7pt);
\draw[thick, MidnightBlue] (-1.15, 0.75) -- (-0.85, 0.75);
\draw[thick, MidnightBlue] (-1.15, 0.65) -- (-0.85, 0.65);

\fill[MidnightBlue] (-1, 0) circle (2pt);
\fill[MidnightBlue] (0.5, 0) circle (2pt);
\draw[ thick, MidnightBlue] (-1, 0) |- (-0.5, -0.7);
\draw[ thick, MidnightBlue] (0.5, 0) |- (0, -0.7);
\draw[ thick, MidnightBlue] (-0.25, -0.7) circle (7pt);
\draw[thick, MidnightBlue] (-0.4, -0.65) -- (-0.1, -0.65);
\draw[thick, MidnightBlue] (-0.4, -0.75) -- (-0.1, -0.75);

\fill[MidnightBlue] (0, 0) circle (2pt);
\fill[MidnightBlue] (1.5, 0) circle (2pt);
\draw[ thick, MidnightBlue] (0, 0) |- (0.5, 0.7);
\draw[ thick, MidnightBlue] (1.5, 0) |- (1, 0.7);
\draw[ thick, MidnightBlue] (0.75, 0.7) circle (7pt);
\draw[thick, MidnightBlue] (0.6, 0.65) -- (0.9, 0.65);
\draw[thick, MidnightBlue] (0.6, 0.75) -- (0.9, 0.75);

\fill[MidnightBlue] (1, 0) circle (2pt);
\draw[thick, MidnightBlue] (1, 0) |- (1.75, -0.7);
\draw[thick, MidnightBlue,  densely dotted] (1.8, -0.7) -- (2, -0.7);

\end{tikzpicture}

%% file: Triad.tex
\begin{tikzpicture}
\tikzset{every node}=[font=\footnotesize\sffamily]

\draw[rounded corners] (-2,-1.25) rectangle (1.5, 1.75);
\node[anchor = north west] at (-2, 1.75) {\textbf{(b) Braided triad}};

\draw[very thick] (-1.75, 0) -- (1.25, 0);

\fill[MidnightBlue] (-1.5, 0) circle (2pt);
\fill[MidnightBlue] (0, 0) circle (2pt);
\draw[ thick, MidnightBlue] (-1.5, 0) |- (-1.25, 0.5);
\draw[ thick, MidnightBlue] (0, 0) |- (-0.75, 0.5);
\draw[ thick, MidnightBlue] (-1, 0.5) circle (7pt);
\draw[thick, MidnightBlue] (-1.15, 0.55) -- (-0.85, 0.55);
\draw[thick, MidnightBlue] (-1.15, 0.45) -- (-0.85, 0.45);

\fill[MidnightBlue] (-1, 0) circle (2pt);
\fill[MidnightBlue] (0.5, 0) circle (2pt);
\draw[ thick, MidnightBlue] (-1, 0) |- (-0.5, -0.7);
\draw[ thick, MidnightBlue] (0.5, 0) |- (0, -0.7);
\draw[ thick, MidnightBlue] (-0.25, -0.7) circle (7pt);
\draw[thick, MidnightBlue] (-0.4, -0.65) -- (-0.1, -0.65);
\draw[thick, MidnightBlue] (-0.4, -0.75) -- (-0.1, -0.75);

\fill[MidnightBlue] (-0.5, 0) circle (2pt);
\fill[MidnightBlue] (1, 0) circle (2pt);
\draw[ thick, MidnightBlue] (-0.5, 0.5-0.1) arc (70:-70:-0.1);
\draw[ thick, MidnightBlue] (-0.5, 0.5+0.08) |- (0.25, 1);
\draw[ thick, MidnightBlue] (-0.5, 0.5-0.09) -- (-0.5, 0);
\draw[ thick, MidnightBlue] (1, 0) |- (0.75, 1);
\draw[ thick, MidnightBlue] (0.5, 1) circle (7pt);
\draw[thick, MidnightBlue] (0.35, 1.05) -- (0.65, 1.05);
\draw[thick, MidnightBlue] (0.35, 0.95) -- (0.65, 0.95);

\end{tikzpicture}

%% file: Matryoshka.tex
\begin{tikzpicture}
\tikzset{every node}=[font=\footnotesize\sffamily]

\draw[rounded corners] (-2,-1.25) rectangle (2,2);
\node[anchor = north west] at (-2, 2) {\textbf{(a) Matryoshka}};

\draw[very thick] (-1.75, 0) -- (-0.2, 0);
\draw[very thick] (0.2, 0) -- (1.75, 0);
\draw[very thick, densely dotted] (-0.1, 0) -- (0.1, 0);

\fill[MidnightBlue] (-1.5, 0) circle (2pt);
\fill[MidnightBlue] (1.5, 0) circle (2pt);
\draw[ thick, MidnightBlue] (-1.5, 0) |- (-0.25, 1);
\draw[ thick, MidnightBlue] (1.5, 0) |- (0.25, 1);
\draw[ thick, MidnightBlue] (0, 1) circle (7pt);
\draw[thick, MidnightBlue] (-0.15, 1.05) -- (0.15, 1.05);
\draw[thick, MidnightBlue] (-0.15, 0.95) -- (0.15, 0.95);

\fill[MidnightBlue] (-1, 0) circle (2pt);
\fill[MidnightBlue] (1, 0) circle (2pt);
\draw[ thick, MidnightBlue] (-1, 0) |- (-0.25, -0.7);
\draw[ thick, MidnightBlue] (1, 0) |- (0.25, -0.7);
\draw[ thick, MidnightBlue] (0, -0.7) circle (7pt);
\draw[thick, MidnightBlue] (-0.15, -0.65) -- (0.15, -0.65);
\draw[thick, MidnightBlue] (-0.15, -0.75) -- (0.15, -0.75);

\fill[MidnightBlue] (-0.5, 0) circle (2pt);
\fill[MidnightBlue] (0.5, 0) circle (2pt);
\draw[ thick, MidnightBlue] (-0.5, 0) |- (-0.25, 0.4);
\draw[ thick, MidnightBlue] (0.5, 0) |- (0.25, 0.4);
\draw[ thick, MidnightBlue] (0, 0.4) circle (7pt);
\draw[thick, MidnightBlue] (-0.15, 0.45) -- (0.15, 0.45);
\draw[thick, MidnightBlue] (-0.15, 0.35) -- (0.15, 0.35);

\end{tikzpicture}

%% file: Enclosed.tex
\begin{tikzpicture}
\tikzset{every node}=[font=\footnotesize\sffamily]

\draw[rounded corners] (-2,-1.25) rectangle (2,2);
\node[anchor = north west] at (-2, 2) {\textbf{(b) Enclosed braided}};

\draw[very thick] (-1.75, 0) -- (1.75, 0);

\fill[MidnightBlue] (-1.5, 0) circle (2pt);
\fill[MidnightBlue] (0, 0) circle (2pt);
\fill[MidnightBlue] (1.5, 0) circle (2pt);
\draw[ thick, MidnightBlue] (-1.5, 0) |- (-0.25, 1);
\draw[ thick, MidnightBlue] (0, 0) -- (0, 0.75);
\draw[ thick, MidnightBlue] (1.5, 0) |- (0.25, 1);
\draw[ thick, MidnightBlue] (0, 1) circle (7pt);
\draw[thick, MidnightBlue] (-0.15, 1.05) -- (0.15, 1.05);
\draw[thick, MidnightBlue] (-0.15, 0.95) -- (0.15, 0.95);

\fill[MidnightBlue] (-0.75, 0) circle (2pt);
\fill[MidnightBlue] (0.75, 0) circle (2pt);
\draw[ thick, MidnightBlue] (-0.75, 0) |- (-0.25, -0.7);
\draw[ thick, MidnightBlue] (0.75, 0) |- (0.25, -0.7);
\draw[ thick, MidnightBlue] (0, -0.7) circle (7pt);
\draw[thick, MidnightBlue] (-0.15, -0.65) -- (0.15, -0.65);
\draw[thick, MidnightBlue] (-0.15, -0.75) -- (0.15, -0.75);

\end{tikzpicture}

%% file: Splitting1.tex
\begin{tikzpicture}
\tikzset{every node}=[font=\footnotesize\sffamily]

\draw[rounded corners] (-2,-1.75) rectangle (1, 1.75);
\node[anchor = north west] at (-2, 1.75) {\textbf{(a)}};

\draw[very thick] (-1.75, 0) -- (0.75, 0);

\fill[MidnightBlue] (-1.5, 0) circle (2pt);
\node at (-1.5, -0.25) {$\gamma$};
\fill[MidnightBlue] (0.5, 0) circle (2pt);
\node at (0.5, -0.25) {$\gamma$};
\draw[ thick, MidnightBlue] (-1.5, 0) |- (-0.75, 1);
\draw[ thick, MidnightBlue] (0.5, 0) |- (-0.25, 1);
\draw[ thick, MidnightBlue] (-0.5, 1) circle (7pt);
\draw[thick, MidnightBlue] (-0.65, 1.05) -- (-0.35, 1.05);
\draw[thick, MidnightBlue] (-0.65, 0.95) -- (-0.35, 0.95);


\fill[MidnightBlue] (-0.5, 0) circle (2pt);
\draw[ thick, MidnightBlue] (-0.5, 0) -- (-0.5, -0.47);
\draw[ thick, MidnightBlue] (-0.5, -0.7) circle (7pt);
\draw[thick, MidnightBlue] (-0.65, -0.65) -- (-0.35, -0.65);
\draw[thick, MidnightBlue] (-0.65, -0.75) -- (-0.35, -0.75);

\draw (-1.75, -1.25) -- (0.75, -1.25);
\draw (-1.5, -1.35) -- (-1.5, -1.15);
\draw (-0.5, -1.35) -- (-0.5, -1.15);
\draw (0.5, -1.35) -- (0.5, -1.15);

\node at (-1, -1.45) {$\phi$};
\node at (0, -1.45) {$\phi$};

\end{tikzpicture}

%% file: Splitting2.tex
\begin{tikzpicture}
\tikzset{every node}=[font=\footnotesize\sffamily]

\draw[rounded corners] (-2,-1.75) rectangle (2,1.75);
\node[anchor = north west] at (-2, 1.75) {\textbf{(b)}};

\draw[very thick] (-1.75, 0) -- (0.9, 0);
\draw[very thick, densely dotted] (0.95, 0) -- (1.3, 0);
\draw[very thick] (1.35, 0) -- (1.75, 0);

\fill[MidnightBlue] (-1.5, 0) circle (2pt);
\node at (-1.5, -0.25) {$\gamma$};
\fill[MidnightBlue] (0, 0) circle (2pt);
\node at (0, -0.25) {$\gamma_1$};
\fill[MidnightBlue] (0.75, 0) circle (2pt);
\node at (0.75, -0.25) {$\gamma_2$};
\fill[MidnightBlue] (1.5, 0) circle (2pt);
\node at (1.5, -0.25) {$\gamma_n$};
\draw[ thick, MidnightBlue] (-1.5, 0) |- (-1, 1);
\draw[ thick, MidnightBlue] (0, 0) |- (0.75, 0.5);
\draw[ thick, MidnightBlue] (0.75, 0) |- (-0.5, 1);
\draw[ thick, MidnightBlue] (1.5, 0) |- (0.75, 0.5);
\draw[ thick, MidnightBlue] (-0.75, 1) circle (7pt);
\draw[thick, MidnightBlue] (-0.9, 1.05) -- (-0.6, 1.05);
\draw[thick, MidnightBlue] (-0.9, 0.95) -- (-0.6, 0.95);


\fill[MidnightBlue] (-0.75, 0) circle (2pt);
\draw[ thick, MidnightBlue] (-0.75, 0) -- (-0.75, -0.47);
\draw[ thick, MidnightBlue] (-0.75, -0.7) circle (7pt);
\draw[thick, MidnightBlue] (-0.9, -0.65) -- (-0.6, -0.65);
\draw[thick, MidnightBlue] (-0.9, -0.75) -- (-0.6, -0.75);

\draw (-1.75, -1.25) -- (1.75, -1.25);
\draw (-1.5, -1.35) -- (-1.5, -1.15);
\draw (-0.75, -1.35) -- (-0.75, -1.15);
\draw (0, -1.35) -- (0, -1.15);
\draw (0.75, -1.35) -- (0.75, -1.15);
\draw (1.5, -1.35) -- (1.5, -1.15);

\node at (-1.125, -1.45) {$\phi$};
\node at (-0.375, -1.45) {$\phi$};
\node at (0.375, -1.45) {$0$};
\node at (1.125, -1.45) {$0$};

\end{tikzpicture}

%% file: TableL.tex
\begin{table*}[t]
\caption{Right and left collapse operators $\COL_R$ and $\COL_L$ for each atomic setup depicted in \figref{fig:setups}. We assume arbitrary phase shifts $\phi_1,\phi_2,\phi_3$ and arbitrary bare relaxation rates $\gamma_{kR}, \gamma_{kL}$ at each coupling point $k=1,2,3,4$.}
\label{tab:L}
\centering
\setlength{\extrarowheight}{8pt}
\setlength{\tabcolsep}{10pt}
\begin{tabular}{c l}
\hline\hline
Topology & Collapse operator \\[4pt]
\hline
Small & 
$\COL_R = \sqrt{\gamma_{1R}}\,e^{i\phi}\sma + \sqrt{\gamma_{2R}}\smb$ \\
& $\COL_L = \sqrt{\gamma_{1L}}\sma + \sqrt{\gamma_{2L}}\,e^{i\phi}\smb$\\
Separate & 
$\COL_R = \left(\sqrt{\gamma_{1R}}\, e^{i(\phi_1+\phi_2+\phi_3)} + \sqrt{\gamma_{2R}}\, e^{i(\phi_2+\phi_3)}\right)\sma + \left(\sqrt{\gamma_{3R}}\, e^{i\phi_3} + \sqrt{\gamma_{4R}}\right)\smb$\\
& $\COL_L = \left(\sqrt{\gamma_{2L}}\, e^{i\phi_1} + \sqrt{\gamma_{1L}}\right)\sma + \left(\sqrt{\gamma_{4L}}\, e^{i(\phi_1+\phi_2+\phi_3)} + \sqrt{\gamma_{3L}}\, e^{i(\phi_1+\phi_2)}\right)\smb$\\
Nested & 
$\COL_R =  \left(\sqrt{\gamma_{1R}}\, e^{i(\phi_1+\phi_2+\phi_3)} + \sqrt{\gamma_{4R}}\right)\sma
+ \left(\sqrt{\gamma_{2R}}\, e^{i(\phi_2+\phi_3)} + \sqrt{\gamma_{3R}}\,e^{i\phi_3}\right)\smb$\\
& $\COL_L = \left(\sqrt{\gamma_{4L}}\, e^{i(\phi_1+\phi_2+\phi_3)} + \sqrt{\gamma_{1L}}\right)\sma + \left(\sqrt{\gamma_{3L}}\, e^{i(\phi_1+\phi_2)} + \sqrt{\gamma_{2L}}\,e^{i\phi_1}\right)\smb$ \\
Braided & 
$\COL_R = \left(\sqrt{\gamma_{1R}}\, e^{i(\phi_1+\phi_2+\phi_3)} + \sqrt{\gamma_{3R}}\, e^{i\phi_3}\right)\sma + \left(\sqrt{\gamma_{2R}}\, e^{i(\phi_2+\phi_3)} + \sqrt{\gamma_{4R}}\right)\smb$\\
& $\COL_L =\left(\sqrt{\gamma_{1L}} + \sqrt{\gamma_{3L}}\, e^{i(\phi_1+\phi_2)}\right)\sma + \left(\sqrt{\gamma_{2L}}\, e^{i\phi_1} + \sqrt{\gamma_{4L}}\,e^{i(\phi_1+\phi_2+\phi_3)}\right)\smb$\\ [12pt]
\hline\hline
\end{tabular} 
\end{table*}

%% file: TableME_full.tex
\begin{table*}[t]
\caption{Frequency shifts, exchange interaction, individual and collective decays [$\delta\omega_j, g, \Gamma_j, \Gammacol$ in \eqref{eq:me}] for small and giant atoms chirally coupled to a 1D open waveguide. We assume arbitrary phase shifts $\phi_1,\phi_2,\phi_3$ and arbitrary bare relaxation rates $\gamma_{kR}, \gamma_{kL}$ at each coupling point $k=1,2,3,4$.}
\label{tab:me_ctes_full}
\centering
\setlength{\extrarowheight}{8pt}
\setlength{\tabcolsep}{10pt}
\begin{tabular}{c c l}
\hline\hline
Coefficient & Topology & Expression for two atoms, $a$ and $b$\\[4pt]\hline
\multirow{8}{*}{\begin{tabular}{c}\textbf{Frequency shifts,}\\ $\delta\omegaA,\, \delta\omegaB$\end{tabular}} & Small  & $0$\\[-6pt]
&& $0$\\
&Separate   & $(\sqrt{\gamma_{1R}\gamma_{2R}}+\sqrt{\gamma_{1L}\gamma_{2L}})\sin\phi_1$\\[-6pt]
&&$(\sqrt{\gamma_{3R}\gamma_{4R}}+\sqrt{\gamma_{3L}\gamma_{4L}})\sin\phi_3$\\
&Nested   & 
$(\sqrt{\gamma_{1R}\gamma_{4R}}+\sqrt{\gamma_{1L}\gamma_{4L}})\sin(\phi_1+\phi_2+\phi_3)$\\[-6pt]
&& $(\sqrt{\gamma_{2R}\gamma_{3R}}+\sqrt{\gamma_{2L}\gamma_{3L}})\sin\phi_2$\\
&Braided   & $(\sqrt{\gamma_{1R}\gamma_{3R}}+\sqrt{\gamma_{1L}\gamma_{3L}})\sin(\phi_1+\phi_2)$ \\[-6pt] 
&& $(\sqrt{\gamma_{2R}\gamma_{4R}}+\sqrt{\gamma_{2L}\gamma_{4L}})\sin(\phi_2+\phi_3)$ \\[4pt]
\hline
\multirow{8}{*}{\begin{tabular}{c}\textbf{Individual decays,}\\ $\GammaA,\, \GammaB$\end{tabular}} & Small  & $\gamma_{1R} + \gamma_{1L}$\\[-6pt]
&&$\gamma_{2R} + \gamma_{2L}$ \\
&Separate   & $\gamma_{1R} + \gamma_{1L} + \gamma_{2R} + \gamma_{2L} + 2(\sqrt{\gamma_{1R}\gamma_{2R}} +\sqrt{\gamma_{1L}\gamma_{2L}}) \cos\phi_1$\\[-6pt]
&&$\gamma_{3R} + \gamma_{3L} + \gamma_{4R} + \gamma_{4L} + 2(\sqrt{\gamma_{3R}\gamma_{4R}} +\sqrt{\gamma_{3L}\gamma_{4L}}) \cos\phi_3$\\
&Nested   & $\gamma_{1R} + \gamma_{1L} + \gamma_{4R} + \gamma_{4L} + 2(\sqrt{\gamma_{1R}\gamma_{4R}} +\sqrt{\gamma_{1L}\gamma_{4L}}) \cos(\phi_1+\phi_2+\phi_3)$\\[-6pt]
&& $\gamma_{2R} + \gamma_{2L} + \gamma_{3R} + \gamma_{3L} + 2(\sqrt{\gamma_{2R}\gamma_{3R}} +\sqrt{\gamma_{2L}\gamma_{3L}}) \cos\phi_2$ \\
&Braided & $\gamma_{1R} + \gamma_{1L} +\gamma_{3R} + \gamma_{3L} + 2(\sqrt{\gamma_{1R}\gamma_{3R}} + \sqrt{\gamma_{1L}\gamma_{3L}}) \cos(\phi_1 + \phi_2)$\\[-6pt]
&& $\gamma_{2R} + \gamma_{2L} +\gamma_{4R} + \gamma_{4L} + 2(\sqrt{\gamma_{2R}\gamma_{4R}} + \sqrt{\gamma_{2L}\gamma_{4L}}) \cos(\phi_2 + \phi_3)$\\[4pt]
\hline
\multirow{8}{*}{\begin{tabular}{c}\textbf{Collective decay,}\\ $\Gammacol$\end{tabular}} & Small  & $\sqrt{\gamma_{1R}\gamma_{2R}}\,e^{i\phi}+\sqrt{\gamma_{1L}\gamma_{2L}}\,e^{-i\phi}$\\
&Separate   & $\sqrt{\gamma_{1R}\gamma_{3R}}\, e^{i(\phi_1+\phi_2)} + \sqrt{\gamma_{1R}\gamma_{4R}}\, e^{i(\phi_1+\phi_2+\phi_3)} + \sqrt{\gamma_{2R}\gamma_{3R}}\, e^{i\phi_2} $\\[-6pt]
&& $+\sqrt{\gamma_{2R}\gamma_{4R}}\,e^{i(\phi_2+\phi_3)}
+\sqrt{\gamma_{1L}\gamma_{3L}}\, e^{-i(\phi_1+\phi_2)}
+ \sqrt{\gamma_{1L}\gamma_{4L}}\, e^{-i(\phi_1+\phi_2+\phi_3)}$\\[-6pt]
&& $+ \sqrt{\gamma_{2L}\gamma_{3L}}\, e^{-i\phi_2} +\sqrt{\gamma_{2L}\gamma_{4L}}\,e^{-i(\phi_2+\phi_3)}$\\
&Nested   & $\sqrt{\gamma_{1R}\gamma_{2R}}\, e^{i\phi_1} 
+ \sqrt{\gamma_{1R}\gamma_{3R}}\, e^{i(\phi_1+\phi_2)} 
+ \sqrt{\gamma_{2R}\gamma_{4R}}\, e^{-i(\phi_2+\phi_3)}$\\[-6pt]
&& $+\sqrt{\gamma_{3R}\gamma_{4R}}\,e^{-i\phi_3} 
+ \sqrt{\gamma_{1L}\gamma_{2L}}\, e^{-i\phi_1} 
+ \sqrt{\gamma_{1L}\gamma_{3L}}\, e^{-i(\phi_1+\phi_2)}$\\[-6pt]
&& $+ \sqrt{\gamma_{2L}\gamma_{4L}}\, e^{i(\phi_2+\phi_3)} +\sqrt{\gamma_{3L}\gamma_{4L}}\,e^{i\phi_3}$\\
&Braided  & $\sqrt{\gamma_{1R}\gamma_{2R}}\, e^{i\phi_1} 
+ \sqrt{\gamma_{1R}\gamma_{4R}}\, e^{i(\phi_1+\phi_2+\phi_3)} 
+ \sqrt{\gamma_{2R}\gamma_{3R}}\, e^{-i\phi_2}$ \\[-6pt]
&& $+\sqrt{\gamma_{3R}\gamma_{4R}}\,e^{i\phi_3} 
+ \sqrt{\gamma_{1L}\gamma_{2L}}\, e^{-i\phi_1} 
+ \sqrt{\gamma_{1L}\gamma_{4L}}\, e^{-i(\phi_1+\phi_2+\phi_3)}$\\[-6pt]
&&$+\sqrt{\gamma_{2L}\gamma_{3L}}\, e^{i\phi_2} +\sqrt{\gamma_{3L}\gamma_{4L}}\,e^{-i\phi_3}$ \\[4pt]
\hline
\multirow{8}{*}{\begin{tabular}{c}\textbf{Exchange}\\[-6pt] \textbf{interaction,}\\ $g$\end{tabular}} & Small  & $[\sqrt{\gamma_{1R}\gamma_{2R}}\,e^{i\phi}-\sqrt{\gamma_{1L}\gamma_{2L}}\,e^{-i\phi}]/2i$\\
& Separate   & $[\sqrt{\gamma_{1R}\gamma_{3R}}\, e^{i(\phi_1+\phi_2)} + \sqrt{\gamma_{1R}\gamma_{4R}}\, e^{i(\phi_1+\phi_2+\phi_3)} + \sqrt{\gamma_{2R}\gamma_{3R}}\, e^{i\phi_2}$ \\[-6pt]
&& $+\sqrt{\gamma_{2R}\gamma_{4R}}\,e^{i(\phi_2+\phi_3)}
-\sqrt{\gamma_{1L}\gamma_{3L}}\, e^{-i(\phi_1+\phi_2)} 
- \sqrt{\gamma_{1L}\gamma_{4L}}\, e^{-i(\phi_1+\phi_2+\phi_3)}$ \\[-6pt] 
&& $- \sqrt{\gamma_{2L}\gamma_{3L}}\, e^{-i\phi_2}
-\sqrt{\gamma_{2L}\gamma_{4L}}\,e^{-i(\phi_2+\phi_3)}]/2i$ \\
&Nested   & $[\sqrt{\gamma_{1R}\gamma_{2R}}\, e^{i\phi_1} 
+ \sqrt{\gamma_{1R}\gamma_{3R}}\, e^{i(\phi_1+\phi_2)} 
- \sqrt{\gamma_{2R}\gamma_{4R}}\, e^{-i(\phi_2+\phi_3)}$\\[-6pt]
&& $-\sqrt{\gamma_{3R}\gamma_{4R}}\,e^{-i\phi_3} 
- \sqrt{\gamma_{1L}\gamma_{2L}}\, e^{-i\phi_1} 
- \sqrt{\gamma_{1L}\gamma_{3L}}\, e^{-i(\phi_1+\phi_2)}$\\[-6pt]
&& $+ \sqrt{\gamma_{2L}\gamma_{4L}}\, e^{i(\phi_2+\phi_3)} +\sqrt{\gamma_{3L}\gamma_{4L}}\,e^{i\phi_3}]/2i$\\
&Braided  & $[\sqrt{\gamma_{1R}\gamma_{2R}}\, e^{i\phi_1} 
+ \sqrt{\gamma_{1R}\gamma_{4R}}\, e^{i(\phi_1+\phi_2+\phi_3)} 
- \sqrt{\gamma_{2R}\gamma_{3R}}\, e^{-i\phi_2}$\\[-6pt]
&& $+\sqrt{\gamma_{3R}\gamma_{4R}}\,e^{i\phi_3} 
- \sqrt{\gamma_{1L}\gamma_{2L}}\, e^{-i\phi_1} 
- \sqrt{\gamma_{1L}\gamma_{4L}}\, e^{-i(\phi_1+\phi_2+\phi_3)}$\\[-6pt]
&& $+\sqrt{\gamma_{2L}\gamma_{3L}}\, e^{i\phi_2} -\sqrt{\gamma_{3L}\gamma_{4L}}\,e^{-i\phi_3}]/2i$\\[4pt]
\hline\hline
\end{tabular}
\end{table*}